\documentclass[pdflatex,sn-mathphys,Numbered,iicol]{sn-jnl}

\usepackage{graphicx}
\usepackage{multirow}
\usepackage{amsmath,amssymb,amsfonts}
\usepackage{amsthm}
\usepackage{mathrsfs}
\usepackage[dvipsnames]{xcolor}
\usepackage{textcomp}
\usepackage{manyfoot}
\usepackage{booktabs}
\usepackage{float}
\usepackage{footnote}
\usepackage{xurl}
\usepackage[font=normal]{subfig}
\usepackage{multirow}
\usepackage{tikz}
\usepackage{wrapfig}
\usepackage[nameinlink,noabbrev]{cleveref}
\Crefname{figure}{Fig.}{Figs.}
\usepackage{changepage}
\usepackage{diagbox}
\usepackage{siunitx}

\usepackage{enumitem}
\setlist[enumerate]{topsep=.5em, partopsep=0pt, itemsep=.5em}
\setlist[itemize]{topsep=.5em, partopsep=0pt, itemsep=.5em}

\begin{document}

\title[Train-Free Segmentation in MRI with Cubical Persistent Homology]{Train-Free Segmentation in MRI \mbox{with Cubical Persistent Homology}}

\author[1]{\fnm{Anton} \sur{François}}

\author[2,3]{\fnm{Raphaël} \sur{Tinarrage}}

\affil[1]{
	\orgdiv{Centre G. Borelli}, 
	\orgname{ENS Paris-Saclay}, 
	\orgaddress{\city{Gif-sur-Yvette}, \country{France}}}

\affil[2]{
	\orgname{IST Austria}, \orgaddress{\city{Klosterneuburg}, \country{Austria}}}

\affil[3]{
	\orgname{EMAp, Fundação Getulio Vargas}, \orgaddress{\city{Rio de Janeiro}, \country{Brazil}}}

%
%

\abstract{
	We investigate a framework for train-free MRI segmentation based on Topological Data Analysis.
	The pipeline proceeds in three steps, first identifying the whole object to segment via automatic thresholding, then detecting a distinctive subset whose topology is known in advance, and finally deducing the various components of the segmentation.
	A key ingredient is the extraction of approximate representative cycles from persistence diagrams, which provides an interpretable link between persistent features and anatomical components.
	To clarify the method's scope, we make the underlying topological and intensity assumptions explicit, quantify when they hold on real data, and analyze typical failure modes.
	We evaluate the approach on glioblastoma and on fetal cortical plate segmentation, with comparisons to unsupervised and deep-learning references.
	By operating without large annotated datasets, the method is well suited to scarce-data settings and provides an interpretable baseline and practical initialization for expert refinement or learning-based pipelines.	
}



\maketitle

%
%

\section{Introduction}

\begin{figure*}[h!]
	\includegraphics[width=\linewidth]{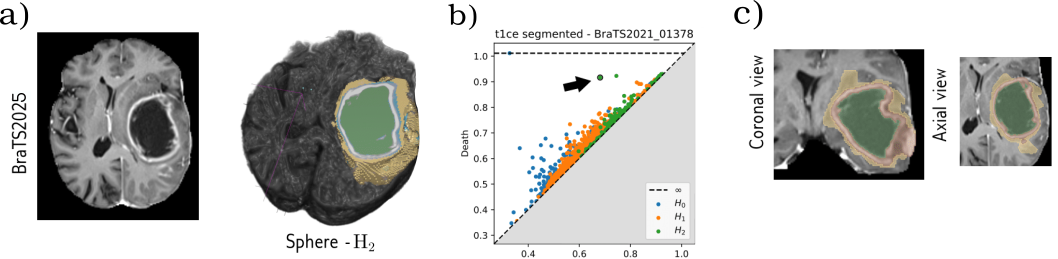}
	\caption{
		\textbf{TDA segmentation overview.} 
		We illustrate a train-free segmentation pipeline based on Topological Data Analysis (TDA) for glioblastoma in BraTS~2025. 
		\textbf{a)}
		We are provided with a 3D MRI and aim at detecting a component of a given topology. 
		\textbf{b)} We automatically select components by analyzing persistence diagrams.
		\textbf{c)}
		Using the strategies detailed in this article, we obtain a segmentation.
		}
	\label{fig:teaser}
\end{figure*}

Anatomical segmentation in Magnetic Resonance Imaging (MRI) refers to the process of identifying and separating different structures within an MRI scan of the body. 
It can be performed by a computer algorithm or a human operator using specialized software.
The algorithm or operator segments the scan into different regions of interest, based on differences in image intensity, shape and size.
Accurate segmentation underpins key clinical tasks, and has inspired extensive methodological development over the last decades \cite{pham2000current,sharma2010automated,litjens2017survey,gao2025medical}.

In biomedical segmentation, substantial effort has focused on \textit{glioblastoma}. It is the most common malignant primary brain tumor in adults, diffuse, variably aggressive, and difficult to prognosticate.
Its segmentation typically involves three regions: the peritumoral Edema (ED), composed of invaded tissue; the Tumorous Core (TC), representing the primary tumor mass; and the Enhancing Tumor (ET), usually the surgical target, along with the necrotic portions of the tumor.
Their union is the Whole Tumor (WT), the entirety of the disease (see \Cref{fig:segmentation_example}).
	
Accurate segmentation of glioblastoma is important for several reasons. 
First, it enables medical professionals to make informed treatment decisions, such as the choice of surgical intervention or radiation therapy, by providing a clear understanding of the size and location of the tumor. 
Second, it is a valuable tool for monitoring disease progression and evaluating the effectiveness of treatment over time. 
Finally, many studies rely on segmentation to extract clinically relevant information \cite{silvestre:hal-04871308, bakas2018identifying}. The segmentation quality can substantially influence the accuracy and efficiency of downstream medical imaging algorithms, making it a critical preprocessing step.

The most popular segmentation methods include U-Net architectures, which achieve excellent scores.
However, they come with known limitations: training requires large annotated datasets that are expensive to build, models are prone to over-fitting, and they are often oblivious to tissue geometry, sometimes yielding anatomically implausible results \cite{de20242024,verma2025brain}.

This motivates continued development of non-deep learning approaches.
Beyond classical unsupervised methods that remain widely used \cite{zhaoAUCseg2021,faragallah2023efficient}, Topological Data Analysis (TDA) is emerging as a promising tool for automatic segmentation. 
In particular, Persistent Homology (PH) encodes the topology of image structures---quantifying connected components, tunnels, and cavities across thresholds---and thus enables the detection of anatomically meaningful components.

Following this line of work, we investigate whether persistent homology can support \emph{train-free} glioblastoma segmentation, without resorting to learned models.
Our contribution is mainly empirical: we assemble a simple segmentation pipeline that combines classical intensity-based operations with topological cues, evaluate it on standard benchmarks, and delineate the conditions under which it performs reliably.
The pipeline is divided into three simple modules (see \Cref{fig:teaser}):
\textbf{(1)}~identification of the whole object to segment;
\textbf{(2)}~detection of a particular subset; and 
\textbf{(3)}~deduction of the other regions.
In particular, we exploit the fact that the enhancing tumor often exhibits a characteristic spherical shape, which provides a discriminative topological signal.	

This framework can be adapted to different segmentation tasks, provided one component has non-trivial homology and the remaining components lie either inside or outside it. 
We illustrate this extension on fetal brain MRI by targeting cortical plate segmentation.

In summary, our main contributions are:
\begin{enumerate}[label=(\roman*)]
	\item 
		The implementation of a TDA-based framework for segmentation on glioblastoma MRI scans that does not require training data.
	\item 
	The study of an explicit topological model that quantifies how frequently this approach is expected to succeed, and a characterization of its failure modes.
	\item 
	Its evaluation on the BraTS 2025 dataset and comparison with well-established methods.
	\item 
		The adaptation of this framework to another segmentation task: the cortical plate in fetal brain MRI.
\end{enumerate}

The remainder of this article is organized as follows. 
Related work is described in \Cref{sec:related_works}, including a short background on TDA.
We describe the topological assumptions that underlie our method in \Cref{sec:model} and the pipeline in \Cref{sec:methods}.
Experimental results are presented and discussed in \Cref{sec:results}.
The framework is adapted to cortical plate segmentation in \Cref{sec:fetal}.
We conclude in \Cref{sec:conclusions}.

The code for this project is available on GitHub\footnote{GitHub repository: \url{https://github.com/antonfrancois/gliomaSegmentation_TDA}}.
It contains tutorials for tumor\footnote{Tutorial on glioblastoma:~\url{https://github.com/antonfrancois/gliomaSegmentation_TDA/tree/main/notebooks/tutorial_brain_segmentation.ipynb}} and cortical plate\footnote{Tutorial on cortical plate:~\url{https://github.com/antonfrancois/gliomaSegmentation_TDA/tree/main/notebooks/tutorial_fetal_segmentation.ipynb}} segmentation on the datasets BraTS\footnote{BraTS: \url{https://www.synapse.org/Synapse:syn64153130}}
and STA\footnote{STA: \url{https://dataverse.harvard.edu/dataset.xhtml?persistentId=doi:10.7910/DVN/WE9JVR}}.

%
%

\section{Related work}\label{sec:related_works}

\subsection{Unsupervised methods}

Early efforts in automatic segmentation employed hand-crafted feature engineering along with traditional machine learning methods. 
This includes standard thresholding methods such as Otsu's, triangle, region growing, k-means, and Gaussian-mixture clustering \cite{otsu1975threshold,AutomaticMeasurementSister,adams1994seeded,mcqueen1967some,dempster1977maximum}.
In parallel, model-based approaches introduced stronger geometric and anatomical priors, such as active contours \cite{kass1988snakes,chan2001active}, graph-cuts \cite{boykov2001interactive,boykov2006graph}, atlas-based registration \cite{warfield2004simultaneous,iglesias2015multi}, and statistical shape models \cite{cootes1995active,cootes1998active}.

Specifically for glioma segmentation, classical non-deep-learning approaches include atlas-based methods, which utilize a representative brain to propagate a segmentation on an unknown one \cite{bauer2010atlas}; decision forests, for instance used by Zikic et al.\ to segment high-grade gliomas using tissue-specific descriptors \cite{zikic2012decision}; and conditional random fields, a type of probabilistic graphical model that can be used to model inter-pixel spatial relationships, as Wu et al.\ applied to tumors \cite{wu2014brain}.

Another family of segmentation methods relies on tree-based representations to encode the structure of the data. 
Among these, the \textit{merge tree} (conceptually close to persistent homology, see \Cref{subsec:tda}) captures the evolution of connected components across threshold levels. Building on this idea, several authors have proposed alternative tree constructions, designed to incorporate additional structural, geometric, or application-specific information, as reviewed in \cite{carlinet2014comparative}. 

For instance, \cite{ballesterTreeShapesImage2003} introduces the \textit{tree of shapes} representation for greyscale images, where shapes are defined as the connected components of upper and lower level sets after hole filling, yielding a self-dual and contrast-invariant hierarchy.
Shapes are ordered by inclusion, forming a multiscale tree structure. 
As another example, in \cite{carrSimplifyingFlexibleIsosurfaces2004}, a \textit{contour tree} is first constructed from the scalar field and then simplified by pruning arcs according to local geometric importance.
Segmentation is finally obtained by selecting a branch. This foundational work is implemented in \texttt{TTK} \cite{tierny2017topology}.

These methods are very effective at capturing complex connected components. However, they generally lack the ability to discriminate which branch of the tree best corresponds to the target structures, unless the associated shape signatures are learned in a statistical manner. In a later work, instead of selecting a segmentation by cutting a predefined hierarchy, the authors propose to reshape the hierarchy itself by exploiting region attributes and extinction values, thereby generating a much broader and more flexible set of meaningful segmentations \cite{xuHierarchicalSegmentationUsing2017}. This approach, however, requires learning a shape space and is therefore inherently statistical.

\subsection{Deep learning}

The rise of GPU processing capabilities has led to a shift in focus towards deep learning for brain tumor segmentation. 
In the first iterations of the BraTS segmentation challenge, these tasks were often addressed via multi-level segmentation approaches. 
This involved first using a simpler method to segment the image and then a deep-learning method to refine the segmentation.
For example, Islam et al. \cite{islam2020detection} designed a four-step multilevel pipeline combining preprocessing, k-means clustering, thresholding with watershed, and morphological refinement. Zhang et al. \cite{zhang20213d} enhanced tumor boundary detection by fusing FLAIR and T2 modalities and training a dense 2D-CNN with novel architectures and loss functions. To address inter-class ambiguity, Liu et al. \cite{liu2021canet} developed a convolutional attention network (CANet).

Driven by the BraTS competition (reviewed in \Cref{subsec:dataset}) and available data, a growing number of deep learning algorithms for tumor segmentation have been developed since 2014. 
The winners of the most recent competition have all employed such networks, which demonstrates their superior performance when more data is available.

Recent BraTS challenges have shown that models can reach segmentation accuracies comparable to expert neuroradiologists. 
The 2019 winning model, a two-stage cascaded U-Net, achieved mean Dice scores\footnote{To compare a predicted segmentation $X$ to a ground truth $Y$, the \textit{Dice score} is defined as the fraction of overlap:
	\[
	\mathrm{Dice}(X,Y) = \frac{2 \# (X \cap Y)}{\# X+\# Y},
	\]
	where $\#X$ and $\#Y$ denote the number of nonzero voxels.}
 of $0.832$ (ET), $0.836$ (TC), and $0.887$ (WT)  \cite{jiang2020twostagecascadedunet}.
In 2023, performance improved with an ensemble trained on a large synthetic dataset, reaching $0.846$ (ET), $0.876$ (TC), and $0.929$ (WT) \cite{ferreira2024wonbrats2023adult}. 
However, comparing these results requires caution: the 2023 training set was three times larger, and while BraTS~2019 annotations were produced by radiologists and corrected by experts, the 2023 annotations were initialized by DL models and then corrected, which may bias results toward higher agreement \cite{kofler2023approachingpeakgroundtruth}.

Although deep learning has become the norm, it still requires large annotated datasets, is computationally expensive to train, and may yield non-interpretable outputs. 
To address this, recent work used TDA to impose topological constraints on predictions, by encouraging anatomically plausible segmentations that preserve connectivity, component counts, and cavity structure.

\subsection{TDA}\label{subsec:tda}

\begin{figure*}[h!]
	\centering
	\includegraphics[width=\linewidth]{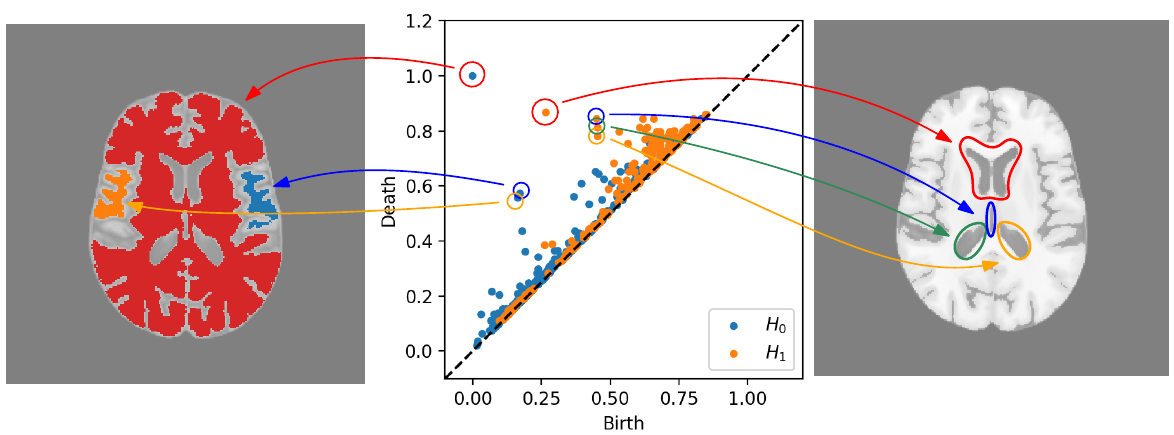}
	\caption{Persistence diagram of the superlevel set filtration of a 2D slice of the SRI template \textbf{(middle)}, representatives of top $H_0$-cycles \textbf{(left)} and top $H_1$-cycles \textbf{(right)}.}
	\label{fig:PD_tuto}
\end{figure*}

\paragraph{Background}

Topological Data Analysis lies at the intersection of computational geometry, algebraic topology, and data analysis. 
It aims to capture relevant geometric and topological information from data \cite{carlsson2009topology,chazal2021introduction}.
Since its emergence in the 2000s, it has been applied to a wide range of problems, including medicine, physics, computer vision, and machine learning \cite{oudot2017persistence,skaf2022topological,singh2023topological}.

We use \textit{cubical persistent homology}, one of the most common tools in TDA.
It is defined as follows.
Given a 3D image $I\colon \Omega\to[0,1]$ and a parameter $t\in [0,1]$, let $I^{\ge t}$ denote the \textit{superlevel set}, that is, the set of voxels of intensity at least $t$:
$$
I^{\ge t} = \{ x \in \Omega \mid I(x)\ge t\}.
$$
It is a topological space---a union of cubes---and we have the relation $I^{\ge s} \subset I^{\ge t}$ whenever $s \ge t$.
Their collection $\{I^{\ge t} \mid t \in [0,1]\}$ forms the \textit{superlevel set filtration} of $I$.

To probe the topology of $I^{\ge t}$, one computes the homology groups $H_i(I^{\ge t})$ (with coefficients in $\mathbb{Z}/2\mathbb{Z}$).
They are vector spaces, whose dimensions characterize the topology of the set.
Roughly, $\dim H_0(I^{\ge t})$ is the number of connected components of $I^{\ge t}$, $\dim H_1(I^{\ge t})$ is the number of ``independent loops'', and $\dim H_2(I^{\ge t})$ is the number of ``independent voids''.
In particular, the sphere has $\dim H_0 =1$, $\dim H_1 =0$, and $\dim H_2 =1$.

Applying the homology functor to the filtration yields a family of vector spaces
$$
\{H_i(I^{\ge t}) \mid t \in [0,1]\}.
$$
Moreover, the inclusions $I^{\ge s} \hookrightarrow I^{\ge t}$ allow one to ``track'' the evolution of the homology. 
That is, one can tell whether an element of $H_i(I^{\ge s})$---called a \textit{cycle}---is still nonzero in $H_i(I^{\ge t})$.

The interval on which a cycle exists is called its \textit{persistence}, and is recorded in a \textit{persistence diagram}.
It is a set of points of the form $p=(t_b,t_d)$, with $t_b \leq t_d$, interpreted as a homological feature born at time $t_b$ and dead at $t_d$.

To each point of the persistence diagram corresponds a \textit{birth voxel}, which gives birth to a cycle (a new connected component, a new $H_1$-cycle, \textit{etc.}), and a \textit{death voxel}, which kills the cycle (merges the component with another one, fills the $H_1$-cycle, \textit{etc.}).
In addition, for each set $I^{\ge t}$ with $t_b \leq t < t_d$, there is a \textit{representative cycle}, that is, a subset that ``identifies'' the cycle.
For $H_0$, the representative cycle is unique, and simply is the connected component of the birth voxel.
However, in higher degrees $H_i$, $i\geq 1$, their identification is a challenging computational task \cite{dey2010optimal,chen2011hardness,escolar2016optimal,obayashi2018volume,dey2019persistent,li2021minimal,cohen2022lexicographic,obayashi2022persistent}.

Rather than computing representative cycles explicitly, we use the connected component of the birth voxel as a proxy.
In our experiments, this simple strategy gave accurate results.

As an illustration, \Cref{fig:PD_tuto} shows the persistence diagram of the superlevel set filtration of a 2D slice (T2 MRI of a healthy brain from the SRI24 Atlas \cite{sri_template}).
The $H_0$-cycles are represented in blue and $H_1$-cycles in orange.
The figure contains three blue dots far from the diagonal. 
They correspond to connected components that evolve independently, without merging into each other.
For each of these persistent cycles, we consider the corresponding point $p=(t_b,t_d)$ of the diagram, extract the pixel of birth, and plot its connected component at time $t_d$. It represents the component just before it merges with another one.
We see that these components are part of the grey matter, disconnected in this slice.
Similarly, four orange dots stand away from the diagonal, with one point particularly off.
In order to represent the corresponding holes, we circled them on the figure, choosing arbitrary representatives.
They correspond to the lateral and third ventricles (again, disconnected in this slice). 

\paragraph{Application to segmentation}

Persistent homology has been used to inject topological structure into segmentation pipelines.
To date, however, only a few approaches operate in an \textit{unsupervised regime}.
These include segmentation of papillary muscles/trabeculae in cardiac imaging \cite{gao2013segmenting,wu2017optimal,chen2017cardiac}, stem-cell clusters in confocal images \cite{nardi2024topology}, and cells/organelles in fluorescence microscopy~\cite{panconi2024three}.

On the other hand, applications of TDA in deep learning for medical imaging can be broadly grouped into three categories: topology-constrained segmentation via  topological losses \cite{clough2019explicit,Clough2020, hu2019topology}, topology-based identification of pathological image components \cite{qaiser2016persistent,qaiser2019fast}, and the use of topological descriptors as features for downstream classification or prediction tasks \cite{rucco2019fast,rucco2020towards,crawford2020predicting}. 

In most existing works, persistence diagrams are treated as global objects for comparison or learning, without exploiting their correspondence with image structures. In contrast, our framework leverages representative cycles to directly associate topological features with anatomically and clinically meaningful regions.

%
%

\section{Topological model}\label{sec:model}

\begin{figure*}[h!]
	\centering
	\includegraphics[width=.99\linewidth]{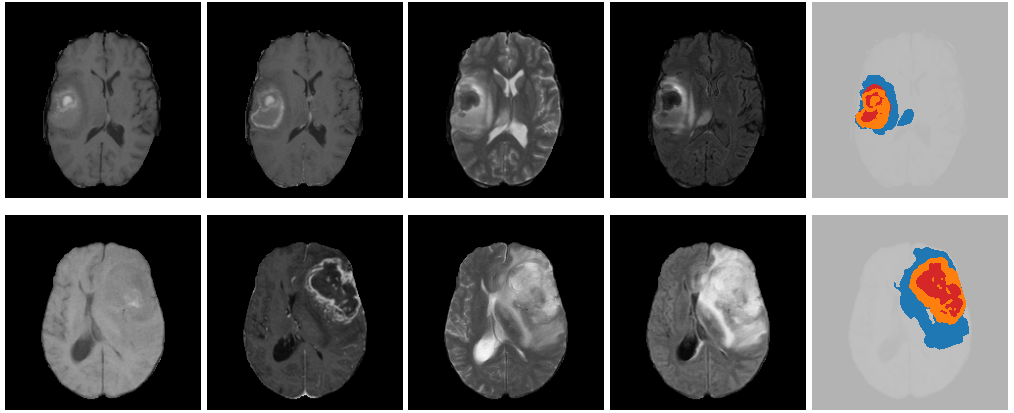}
	\caption{
		\textbf{Glioblastoma segmentation in BraTS 2025.}
		Rows contain horizontal MRI slices of a patient (modalities T1, T1ce, T2, FLAIR) and the provided segmentation (TC: red, ED: blue, ET: orange).
	}
	\label{fig:segmentation_example}
\end{figure*}

We start by outlining the topological insights that support our approach.
It leverages two complementary cues for train-free segmentation: \emph{intensity} and \emph{topology}.
Intensity is used to generate candidate foreground regions, since hyper-intense structures appear in superlevel sets.
Topology is then used to disambiguate these candidates by selecting features in the persistence diagram whose homology matches the expected target.
In \Cref{subsec:definition_model}, we state the conditions under which this strategy is expected to succeed, anticipating the pipeline of \Cref{sec:methods}.
In \Cref{subsec:validation_model}, we provide a quantitative version of these conditions adapted to BraTS.

\subsection{Definition of the model}\label{subsec:definition_model}

Our method can be applied provided there exists a component in the segmentation, referred to as the \emph{geometric object}, whose homology is non-trivial and which separates the other components.
Additionally, the union of all the components, called the \textit{whole object}, must appear hypo- or hyper-intense in the image.
More precisely, we consider the following hypotheses on the ground-truth segmentation:
\begin{enumerate}[label=\textbf{(H\arabic*)},ref=(H\arabic*)]
	\item\label{ch4:enum:global} The whole object is connected, has no holes, and appears significantly hyper-intense in at least one image modality.
	\item\label{ch4:enum:ET} The geometric object has non-trivial homology 
	and appears particularly hyper-intense in at least one image modality.
	\item\label{ch4:enum:TC} Other components lie inside or outside the geometric object.
\end{enumerate}

In the context of the BraTS database, the whole object is referred to as WT (Whole Tumor), and the role of the geometric object is played by the ET (Enhancing Tumor).
Although not systematically satisfied, we found that these components regularly exhibit common characteristics, gathered in \Cref{tab:features_brats}.
An example is given in the second row of \Cref{fig:segmentation_example}.
In the FLAIR modality (penultimate column), WT is associated with the brightest voxels. In T1ce (second column), ET forms a sphere.

\begin{table}[h!]
	\centering
	\begin{tabular}{l|llll}
		& \textbf{WT} & \textbf{TC} & \textbf{ET} & \textbf{ED} \\ \hline
		\textbf{FLAIR} &
		\begin{tabular}[c]{@{}l@{}}hyper\end{tabular} &
		\begin{tabular}[c]{@{}l@{}}hyper\end{tabular} &
		\begin{tabular}[c]{@{}l@{}}hyper\end{tabular} &
		\begin{tabular}[c]{@{}l@{}}hyper\end{tabular} \\ \hline
		\textbf{T1ce} &
		\begin{tabular}[c]{@{}l@{}}--\end{tabular} &
		\begin{tabular}[c]{@{}l@{}}hypo\end{tabular} &
		\begin{tabular}[c]{@{}l@{}}hyper\end{tabular} &
		\begin{tabular}[c]{@{}l@{}}hypo\end{tabular} \\ \hline
		\textbf{Topology} &
		\begin{tabular}[c]{@{}l@{}}connected\end{tabular} &
		\begin{tabular}[c]{@{}l@{}}--\end{tabular} &
		\begin{tabular}[c]{@{}l@{}}spherical\end{tabular} &
		\begin{tabular}[c]{@{}l@{}}--\end{tabular}
	\end{tabular}
	\caption{BraTS: typical intensity (hyper- or hypo-intense) in FLAIR and T1ce for each component, and the expected topology.}
	\label{tab:features_brats}
\end{table}

\subsection{Validation of the model}\label{subsec:validation_model}

To quantify the validity of the model presented above, we formalize the hypotheses as follows:
\begin{itemize}
	\item[\textbf{(H1')}] 
	The median intensity of WT in FLAIR is at least $3/2$ times greater than that of the remaining nonzero voxels in the image.
	\item[\textbf{(H2')}] 
	The median intensity of ET in T1ce is at least $3/2$ times greater than the median intensity of the remaining voxels of WT.
	\item[\textbf{(H3')}]
	After applying two binary dilations to ET and separating its complement in connected components, at least $90\%$ of TC lies inside.
\end{itemize}

We point out that the value of 3/2 in (H1') and (H2') is arbitrary, chosen to represent a significant change in intensity.
Besides, the two binary dilations in (H3') were chosen so as to avoid the effects of ``thin-edged'' tumors that form an open sphere (see the discussion in \Cref{subsec:step2}).

We computed that the assumptions hold on 441 of the 1251 images (35.3\%).
Consistently, we obtain substantially better scores on this subset, as we will present in \Cref{subsec:scores}.

When the assumptions fail---for instance due to weak enhancement or a non-separating enhancing ET---the persistence-based selection may target an incorrect structure and the segmentation quality degrades.
We further analyze such failure modes in \Cref{subsec:qualitative_eval}.
	This highlights a limitation of our approach: it is expected to succeed only in cases where the intensity and topological assumptions are approximately satisfied.

More broadly, MRI data are subject to substantial variability across scanners and patients (contrast changes, noise levels) as well as to intensity inhomogeneities (bias fields related to $B_0$ and $B_1$ effects). 
Such effects can directly impact the validity of hypotheses (H1')--(H3') by reducing intensity contrast and causing the structures to appear fragmented.
In BraTS, described in \Cref{subsec:dataset}, this issue is mitigated by the standardized preprocessing of the dataset.

%
%

\section{Methods}\label{sec:methods}

We now describe our glioma segmentation algorithm.
It is split into three steps, detailed below.

\begin{enumerate}[label={},ref=Module~\arabic*,leftmargin=0cm]
	\item\label{item:step1}
	\textbf{Module 1:} 
	\textbf{Identification of WT} (see \Cref{subsec:step1})
	By \ref{ch4:enum:global}, the FLAIR modality shows a hyper-intense area, corresponding to the whole tumor.
	We extract a candidate from a superlevel set, using a traditional thresholding method.
	
	\item\label{item:step2} 
	\textbf{Module 2:} 
	\textbf{Detection of ET} (see \Cref{subsec:step2})
	Following \ref{ch4:enum:ET}, the enhancing tumor is hyper-intense in T1ce and is expected to exhibit a spherical geometry.
	We compute persistent homology on T1ce restricted to WT, and select a prominent $H_2$ feature, which guides the extraction of ET.

	\item\label{item:step3} 
	\textbf{Module 3:} 
	\textbf{Deduction of TC and ED} (see \Cref{subsec:step3})
	Finally, we classify the components of WT minus ET as being \emph{inside} or \emph{outside} of ET, as suggested by \ref{ch4:enum:TC}, yielding TC and ED.
\end{enumerate}

\subsection{\ref{item:step1}: Identification of WT}\label{subsec:step1}

In this first step, we extract the largest hyper-intense region in the FLAIR image, assumed to correspond to WT.
To do so, we consider the FLAIR image $I_\mathrm{FLAIR}\colon \Omega\rightarrow [0,1]$ and its superlevel set filtration $\{I_\mathrm{FLAIR}^{\geq t} \mid t \in [0,1]\}$.
While the number of voxels increases steadily, we anticipate a sharp increase, precisely when the voxels corresponding to the white and grey matter get included. 
By selecting $t$ just before this event, we obtain a reliable estimate of WT.

More precisely, we perform a simple global binarization followed by largest-component selection.
Let $h$ be the histogram of intensities, computed on a fixed number of bins (equal to 1000 in our experiments).
In addition, we discard the voxels of low intensity (lower than 0.1).
Let $m$ be the mean number of voxels in each bin.
We define $b^*$ as the first bin, starting from $t=1$, whose count is greater than $m$, and let $t^*$ be its intensity value.
Finally, we set WT to be the largest connected component (the component containing the most voxels) of the superlevel set $I_\mathrm{FLAIR}^{\ge t^*}$.

\begin{figure*}[h!]
	\captionsetup[subfloat]{farskip=0pt,captionskip=0.15cm}
	\centering
	\subfloat[Automatic threshold selection]{
		\includegraphics[width=0.49\linewidth]{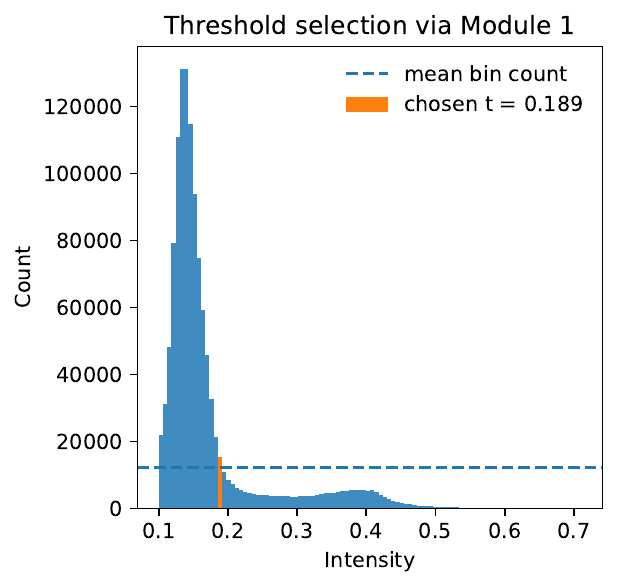}
		}
	\hfill
		\subfloat[Segmentation of WT]{
	\includegraphics[width=0.43\linewidth]{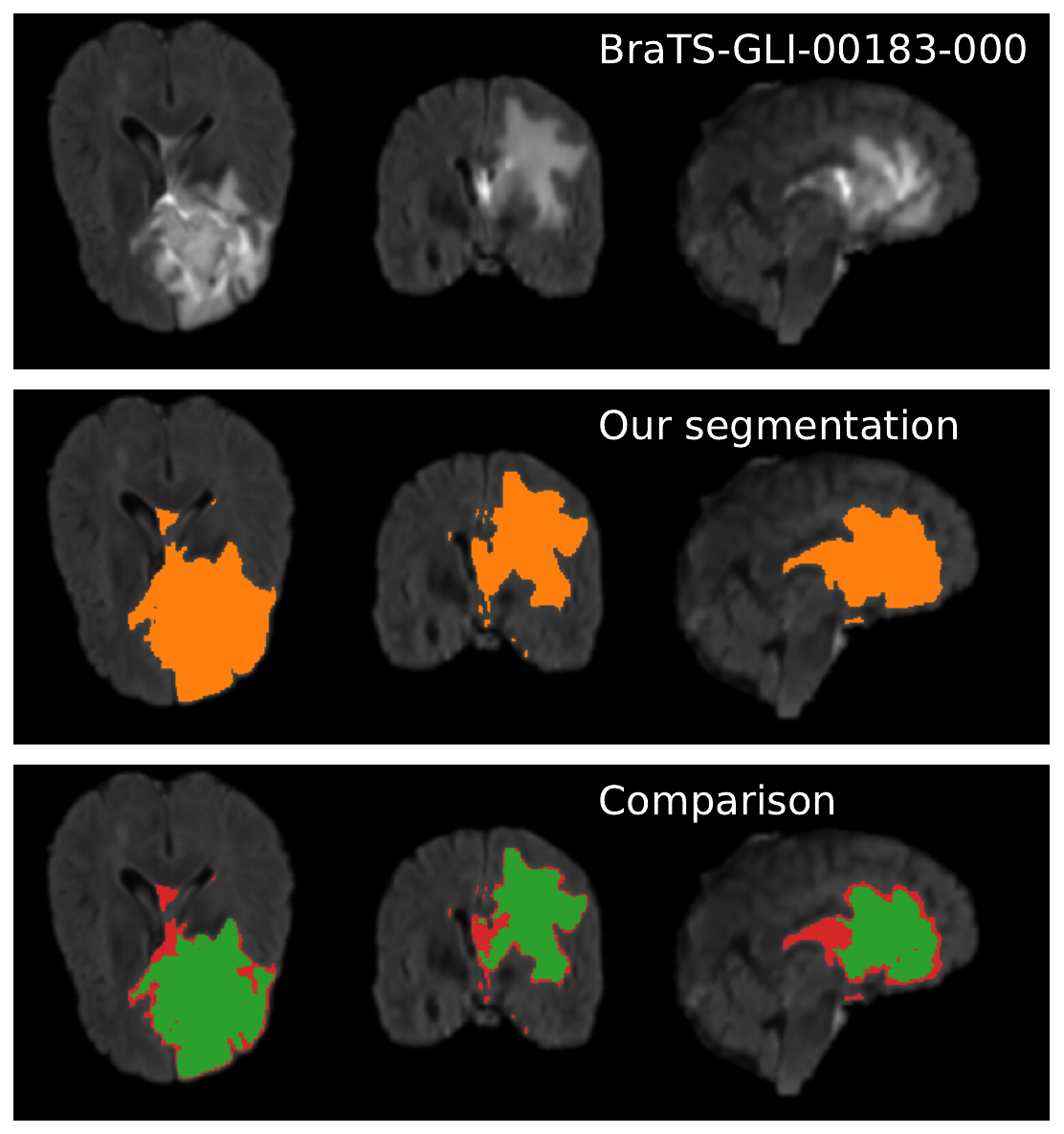}
	}
	\caption{\textbf{Identifying WT via \ref{item:step1}.}  
		We consider a FLAIR-modality MRI in BraTS 2025.
		\textbf{(a)}~Histogram of the image and optimal threshold found by our procedure.
		\textbf{(b)}~Raw image (top), our segmentation of WT in orange (middle), and overlay of the output with the ground truth (green: correctly segmented, orange: mislabeled, red: false positive or negative).
		Dice: 0.888.}
	\label{fig:ch4:suggest_t}
\end{figure*}

This procedure has the advantage of being essentially parameter-free.
It is illustrated in \Cref{fig:ch4:suggest_t}, which shows the intensity histogram, the selected threshold, and the resulting segmentation.
In this example, the obtained component is close to the ground truth, with a Dice score of $0.888$.
This initial segmentation serves as a coarse but robust estimate that is subsequently refined by the geometric and topological components of our pipeline. 

\begin{figure}[ht]
	\centering
	\subfloat[Dice distribution across subjects.\label{fig:boxplot_threshold_methods}]{
		\includegraphics[width=\linewidth]{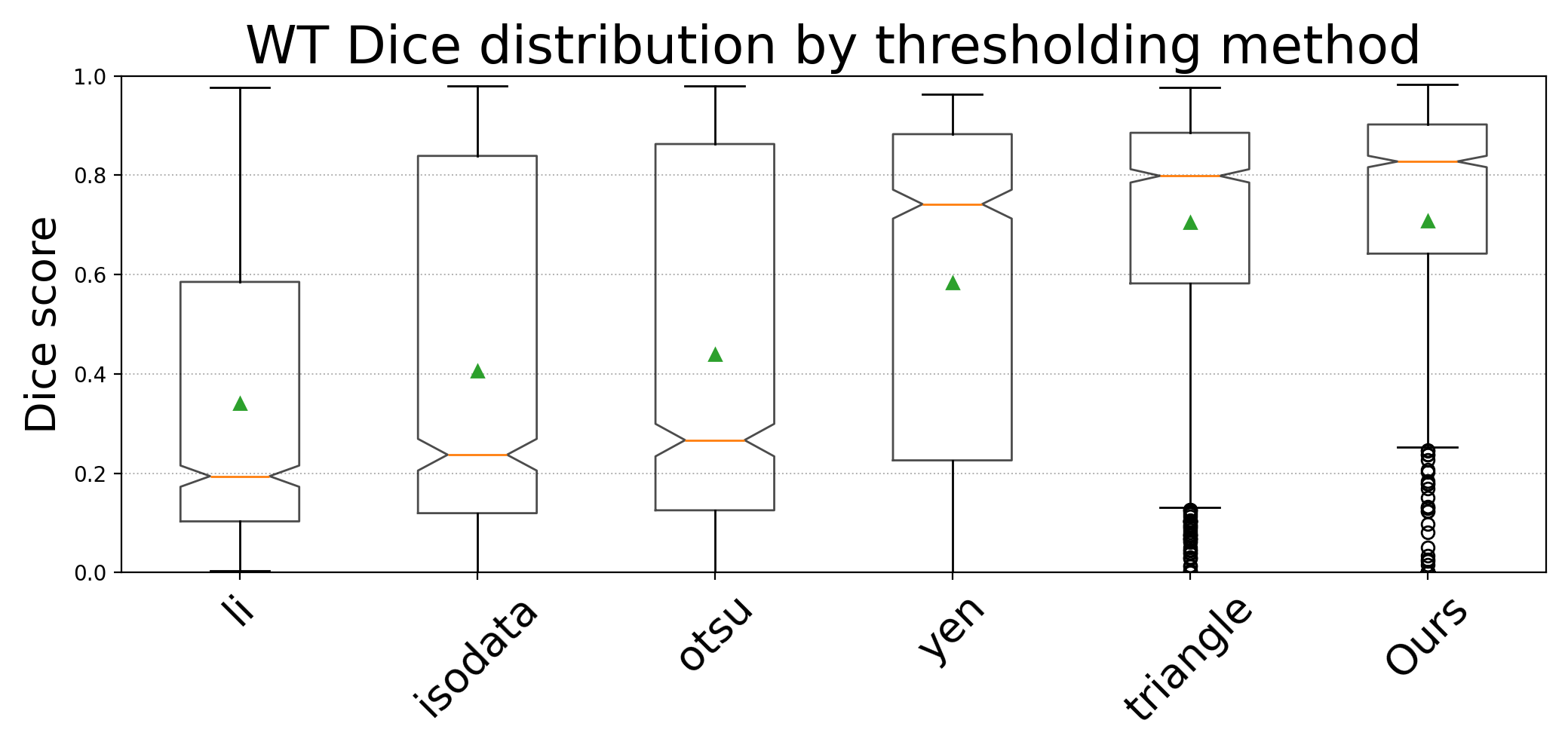}
	}
	\vspace{0.01em}
	
	\subfloat[Mean and median Dice scores\label{tab:threshold_comparison}]{
		\begin{tabular}{lcc}
			\hline
			Method & Mean Dice ($\pm$ std) & Median Dice \\
			\hline
			Li        & $0.341 \pm 0.320$ & $0.194$ \\
			Isodata   & $0.406 \pm 0.345$ & $0.237$ \\
			Otsu      & $0.440 \pm 0.354$ & $0.267$ \\
			Yen       & $0.583 \pm 0.335$ & $0.742$ \\
			Triangle  & $0.705 \pm 0.239$ & $0.799$ \\
			Ours      & $\mathbf{0.708 \pm 0.288}$ & $\mathbf{0.828}$ \\
			\hline
		\end{tabular}
	}
	
	\caption{\textbf{Comparison of thresholding methods.}
			Thresholds were estimated from truncated histograms (intensity $>0.1$).}
	\label{fig:threshold_global}
\end{figure}

\paragraph{Comparison with other methods}

We compare our method with several classical binarization techniques implemented in the \texttt{scikit-image} library, namely Li, IsoData, Otsu, Yen and triangle. 
We discarded mean and minimum methods because of poor performance.
The same preprocessing pipeline was applied to all methods to ensure a fair comparison. 
Thresholds were computed from a truncated intensity histogram, restricted to voxels with intensity greater than 0.1. This truncation reduces the influence of low-intensity voxels and stabilizes the estimation of global thresholds by focusing on the intensity range relevant to the target structure. The truncation threshold was selected empirically to maximize the average Dice score across subjects.

As shown in \Cref{fig:boxplot_threshold_methods,tab:threshold_comparison}, our approach consistently outperforms other classical thresholding methods, even in its simplest form and without any additional refinement.
Interestingly, our procedure shares conceptual similarities with the \textit{triangle} thresholding method introduced in \cite{AutomaticMeasurementSister}. 
That method assumes the presence of a dominant peak near one end of the histogram and searches for a threshold toward the opposite end. Given a normalized histogram, the threshold is defined as the point that maximizes the distance between the histogram and the straight line connecting the main peak to the lowest non-zero bin. 
However, unlike the triangle method, our procedure does not rely on assumptions about the histogram shape, which allows it to remain effective across a wider range of intensity distributions. 

\paragraph{Refinement}

In our experiments, we ultimately adopted a simple preprocessing pipeline: 0--1 normalization and a mild Gaussian smoothing.
Although we extensively tested a range of alternatives (local and patch-wise normalization, histogram equalization, denoising methods such as SUSAN and Non-Local Means, and contrast enhancement), these did not improve performance and sometimes degraded it, likely due to substantial inter-subject variability.

After selecting a threshold $t$, the natural choice is to keep the largest connected component of the thresholded FLAIR image.
In practice, however, the largest component can be dominated by non-tumor hyperintensities or acquisition-related artifacts that produce diffuse regions.
To increase robustness, we therefore prefer a \emph{shape-based} selection: among components above a minimum size (10,000 voxels, i.e., 0.14\% of the volume), we select the most spherical one, measured by voxelized sphericity \cite{wadell1935volume}:
\[
\Phi = \pi^{1/3}\,(6V)^{2/3}/A,
\]
where $V$ is its volume and $A$ its surface area (exposed voxels). 
This heuristic reduces sensitivity to large spurious components while preserving the expected compact morphology of WT.

In addition, we let the threshold $t$ vary in a small window (of a fixed radius equal to 0.02) and select the value corresponding to the most spherical component, quantified by the quantity $\Phi$ above.
In the example in \Cref{fig:ch4:suggest_t}, this fine-tuning improves the Dice score from 0.888 to 0.926.

Lastly, we apply a postprocessing step that fills holes in the segmentation, as suggested by \ref{ch4:enum:global} of our model.
We compute connected components of $\Omega\setminus \mathrm{WT}$, discard the background component, and add the remaining components to WT.
From a biomedical viewpoint, we observed that the presence of holes in WT is often caused by the necrosis, which occasionally appears darker 
on FLAIR.

\paragraph{Evaluation of threshold selection accuracy}

To quantify how close the automatically selected threshold is to an optimal one within this framework, we compare it to an oracle threshold.
For each superlevel set, we extract the largest connected component and compute its Dice score against the ground truth; the threshold yielding the highest score is retained as the reference.

For the FLAIR segmentation obtained with our method (including refinements), the Dice score achieves a mean of $0.76 \pm 0.27$ (std) on the entire dataset. In contrast, the oracle best Dice score reaches a mean of $0.83 \pm 0.18$. 
These results provide an upper bound on the achievable performance within the considered thresholding framework.

\subsection{\ref{item:step2}: Detection of ET}\label{subsec:step2}

\begin{figure*}[h!]
	\captionsetup[subfloat]{farskip=0pt,captionskip=0.15cm}
	\centering
	\subfloat[Automatic cycle selection]{
		\includegraphics[width=0.42\linewidth]{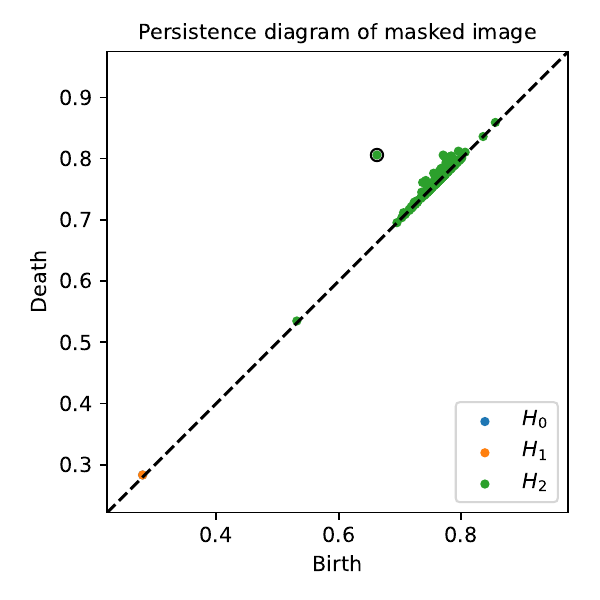}
	}
	\hfill
	\subfloat[Segmentation of ET]{
		\includegraphics[width=0.43\linewidth]{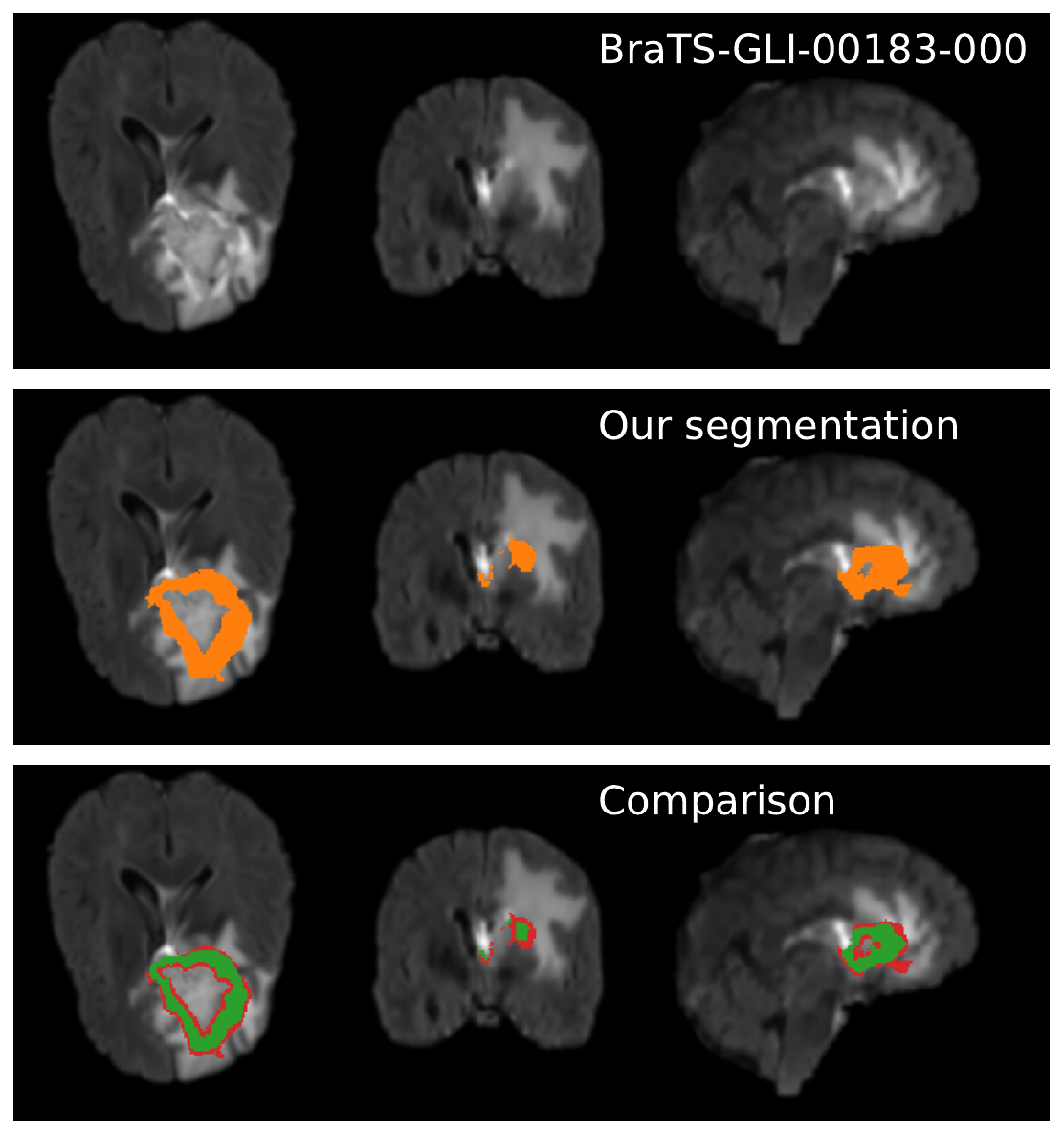}
	}
	\caption{\textbf{Identifying ET via \ref{item:step2}.}
	We consider a T1ce-modality MRI in BraTS 2025 (same patient as \Cref{fig:ch4:suggest_t}).
	\textbf{(a)}~Persistence diagram of T1ce restricted to WT (top $H_2$-cycle circled).
		\textbf{(b)}~Raw image (top), segmentation of ET via \ref{item:step2} in orange (middle), and overlay of the output with the ground truth (green: correctly segmented, orange: mislabeled, red: false positive or negative). 
	Dice: 0.894.}
	\label{fig:segmentation_step2}
\end{figure*}

We now use the T1ce modality together with the previously computed WT to obtain the enhancing tumor ET.
According to \ref{ch4:enum:ET}, this component is the boundary of the tumorous core and is highly intense in T1ce. 
Hence, in the superlevel sets $\{I_\mathrm{T1ce}^{\geq t} \mid t \in [0,1]\}$, we expect to see a sphere, represented by a cycle in the $H_2$-persistence diagram.

This procedure is implemented as follows.
First, we compute the persistent homology of the superlevel set filtration of the image restricted to WT.
Then, we select the $H_2$-feature of highest persistence, that is, the point $(t_b,t_d)$ of the diagram that maximizes $|t_d-t_b|$.
Let $x_b\in\Omega$ be the voxel that gave birth to it.
Following our strategy outlined in \Cref{subsec:tda}, we define ET as the connected component of $x_b$ in the binary image $I_\mathrm{T1ce}^{\ge t_b}$.
This component may not itself be a representative cycle of the homology class; it only contains one.

\Cref{fig:segmentation_step2} shows a concrete example.
On the diagram, one green point appears particularly far from the diagonal: it is the persistent cycle we are looking for.
The resulting segmentation in panel~(b) reaches a Dice score of $0.894$.

\paragraph{Refinement}

In practice, the ground-truth ET may appear thin, weakly contrasted, or even not spherical in T1ce, which can hinder the emergence of an $H_2$ feature.
This is a common challenge in persistent-homology-based pipelines.
To promote robustness of the cycle, we apply a light preprocessing prior to computing persistent homology.
Again, we investigated various combinations of methodological choices and associated parameters.
Specifically, we \textbf{(i)} smooth the image with a Gaussian filter of standard deviation $\sigma=1$, \textbf{(ii)} enhance contrast via \texttt{skimage.filters.rank.enhance\_contrast} with radius $1$, and \textbf{(iii)} apply a greyscale dilation of radius $2$ using \texttt{skimage.morphology.dilation}.
This combination reduces noise and thickens bright structures, thereby stabilizing the persistence diagram and the extraction of ET.

\subsection{\ref{item:step3}: Deduction of the other components}\label{subsec:step3}

We finally identify the components TC and ED. 
This last step does not depend on the initial MRI, but only on WT and ET estimated previously.

Following \ref{ch4:enum:TC}, TC corresponds to the part of the tumor that lies \textit{inside} ET, and ED to the part that lies \textit{outside} ET (and still within WT).
In order to identify them, we consider the subset $\Omega \setminus \mathrm{ET}$---the complement of ET---and compute its connected components.
Note that it may have more than two components.
The outer component is identified as that containing the background, and its restriction to WT is saved in ED.
The others are considered inner and are added to TC.

\begin{figure}[h!]
	\centering
	\includegraphics[width=0.9\linewidth]{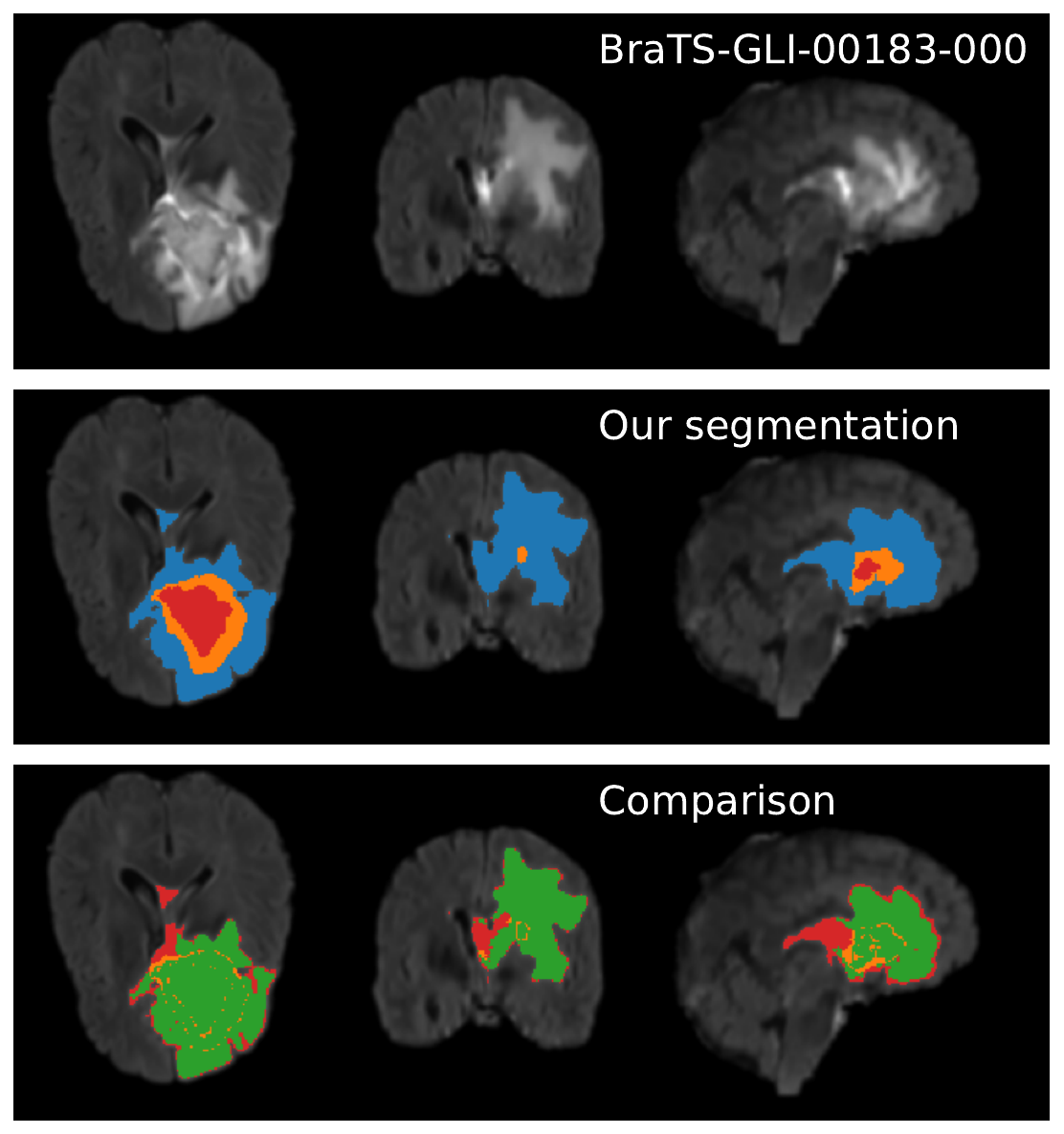}
	\caption{\textbf{Deducing TC and ED via \ref{item:step3}.}  
		These subsets are identified based on their location inside or outside ET. 
		TC: red, ED: blue, ET: orange.
		Dice: 0.960 (TC), 0.884 (ED), 0.894 (ET).
	}
	\label{fig:segmentation_step3}
\end{figure}

The final segmentation is visualized in \Cref{fig:segmentation_step3}.
This is a successful case, achieving Dice scores of 0.960 (TC), 0.884 (ED), and 0.894 (ET).

\paragraph{Running time}

We report the running time of each step of the pipeline, averaged over all volumes of the BraTS collection (mean $\pm$ standard deviation in seconds), computed on the CPU of a personal laptop (Intel Core Ultra 5 125Ux14, 32GB RAM). Preprocessing: $1.05 \pm 0.13$s, \ref{item:step1}: $11.79 \pm 1.20$s, \ref{item:step2}: $10.01 \pm 1.04$s, and \ref{item:step3}: $1.02 \pm 0.18$s. Overall, the complete pipeline runs in about $23.87$s per scan on average, with the first two modules accounting for most of the computation time.

For the computation of cubical persistent homology in \ref{item:step2}, we tested \texttt{giotto-tda}, \texttt{Cubical Ripser}, and \texttt{GUDHI} \cite{tauzin2021giotto,kaji2020cubical,maria2014gudhi}. In our experiments, \texttt{Cubical Ripser} ran the fastest, and we therefore adopted it in the final implementation.

%
%

\section{Results and discussion}\label{sec:results}

\begin{figure*}[h!]
	\centering
	\includegraphics[width=\linewidth]{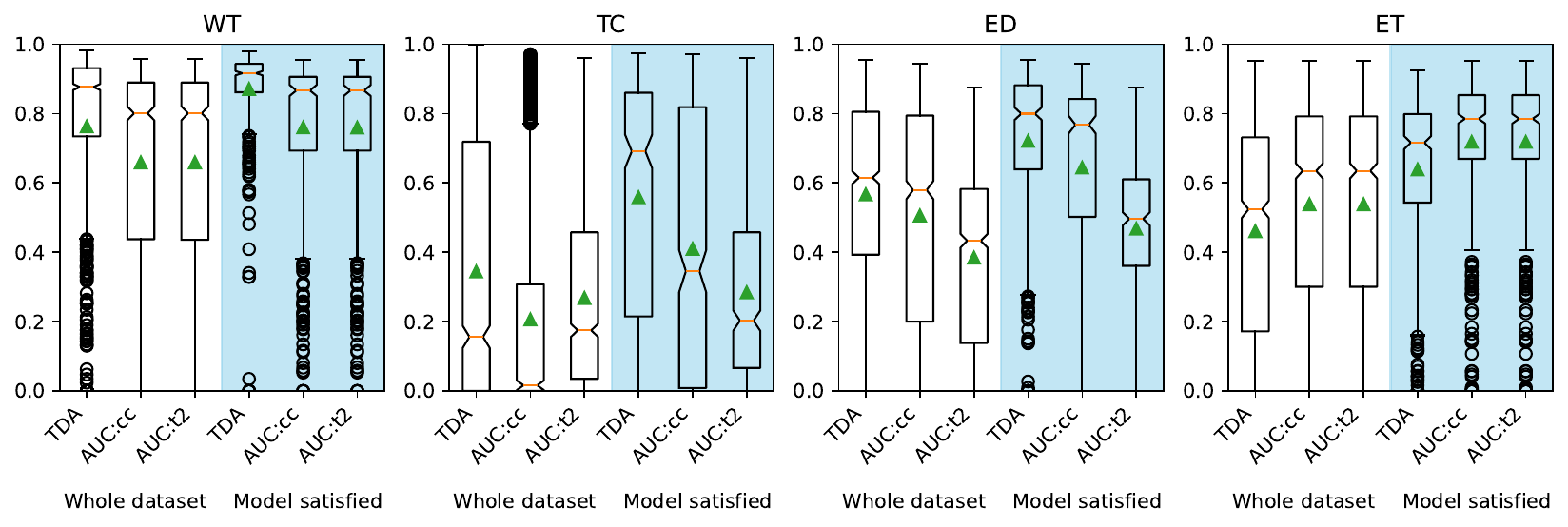}
	\caption{
		\textbf{Results of our method on BraTS 2025.}
		Boxplots of the Dice coefficients on segmentations of the BraTS 2025 dataset, for the four regions (WT, TC, ED and ET), in two scenarios: 
		for the whole dataset (1251 MRIs) comparing our method (\textbf{TDA}) and AUCseg (variations \textbf{AUCseg:CC} and \textbf{AUCseg:T2}, see \Cref{subsubsec:brain_comparison}), and on the subset where the model is valid (441 MRIs), again comparing the methods.}
	\label{fig:benchmark_all}
\end{figure*}

\begin{table*}[h!]
	\centering
	\begin{tabular}{l|ccc}
		\toprule
		\diagbox{\textbf{Region}}{\textbf{Method}} & ~~~~~~\textbf{TDA}~~~~~~ & \textbf{AUCseg:CC} & \textbf{AUCseg:T2} \\
		\midrule
		Whole Tumor (WT)     & \textbf{0.76}$\pm$\textbf{0.27} & $0.66\pm0.29$ & $0.66\pm0.29$ \\
		Tumor Core (TC)      & \textbf{0.34}$\pm$\textbf{0.36} & $0.21\pm0.32$ & $0.27\pm0.27$ \\
		Edema (ED)           & \textbf{0.57}$\pm$\textbf{0.27} & $0.51\pm0.31$ & $0.38\pm0.25$ \\
		Enhancing Tumor (ET) & $0.46\pm0.30$ & \textbf{0.54}$\pm$\textbf{0.30} & \textbf{0.54}$\pm$\textbf{0.30} \\
		\bottomrule
	\end{tabular}
	\caption{
		\textbf{Comparison with another method on BraTS 2025.}
		Mean Dice score on the BraTS 2025 dataset obtained by our method \textbf{TDA} and AUCseg \cite{zhaoAUCseg2021}, following the two pipelines proposed (\textbf{AUCseg:CC} and \textbf{AUCseg:T2}).
		We indicate in bold the highest score(s) for each category.
	}	
	\label{table:brains_comparison}
\end{table*}

In this section, we evaluate our pipeline on the BraTS 2025 GLI-PRE cohort, reporting performance on the full dataset and on the subset where the assumptions from \Cref{sec:model} hold.

\subsection{Dataset}\label{subsec:dataset}

The \textit{Center for Biomedical Image Computing \& Analytics} of the Perelman School of Medicine has run the \textit{Brain Tumor Segmentation} (BraTS) Challenge for a decade. We use GLI-PRE, a subset of BraTS 2025~\cite{de20242024} that also appears in BraTS 2021 \cite{brats_2021,BRATS2021,Bakas2017}; see \cite{bonato2025advancing} for a complete overview of BraTS.

The dataset consists of 1251 MRIs with a spatial resolution of $182 \times 218 \times 182$ voxels, coming in four modalities: Native (T1), contrast-enhanced (T1ce), T2-weighted (T2) and T2 Fluid-Attenuated Inversion Recovery (FLAIR), along with a ``ground-truth'' segmentation for each patient. 
All imaging volumes were acquired using heterogeneous clinical protocols and scanners across multiple contributing institutions. The reference dataset was generated following a preprocessing pipeline that included co-registration to a common anatomical template, resampling to an isotropic resolution of \SI{1}{mm^3}, and skull stripping. These segmentations were made manually and subsequently validated by neuroradiologists.

Although the results of the 2025 challenge have not yet been published, the state-of-the-art methods of the preceding edition are all U-Net-based \cite{luu2022extending,yuan2022evaluating,futrega2022optimized,siddiquee2021redundancy,ma2022nnunet,kotowski2022coupling,ren2022ensemble,jia2022hnf}.
That said, unsupervised methods continue to be proposed, such as AUCseg \cite{zhaoAUCseg2021}, against which we will compare our method in \Cref{subsubsec:brain_comparison}.

\subsection{Scores}\label{subsec:scores}

We applied our algorithm, described in \Cref{sec:methods}, to the entire GLI-PRE dataset.
For each scan, we computed the class-wise Dice coefficients (WT, TC, ED, and ET) against the expert-provided segmentations.
The outcomes are presented in \Cref{fig:benchmark_all} and \Cref{table:brains_comparison}, alongside the performance of AUCseg (further analyzed in \Cref{subsubsec:brain_comparison}).

The boxplots show that our method performs well for WT (mean value and standard deviation of $0.76\pm0.27$).
Performance is moderate on ED and ET (respectively $0.57\pm0.27$ and $0.46\pm0.30$), with the lowest score for TC ($0.34\pm0.36$).

\begin{figure*}[h!]
	\captionsetup[subfloat]{farskip=0pt,captionskip=0.15cm}
	\centering
	\subfloat[Cases where the model is valid]{
		\begin{minipage}{.985\linewidth}
			\includegraphics[width=.329\linewidth]{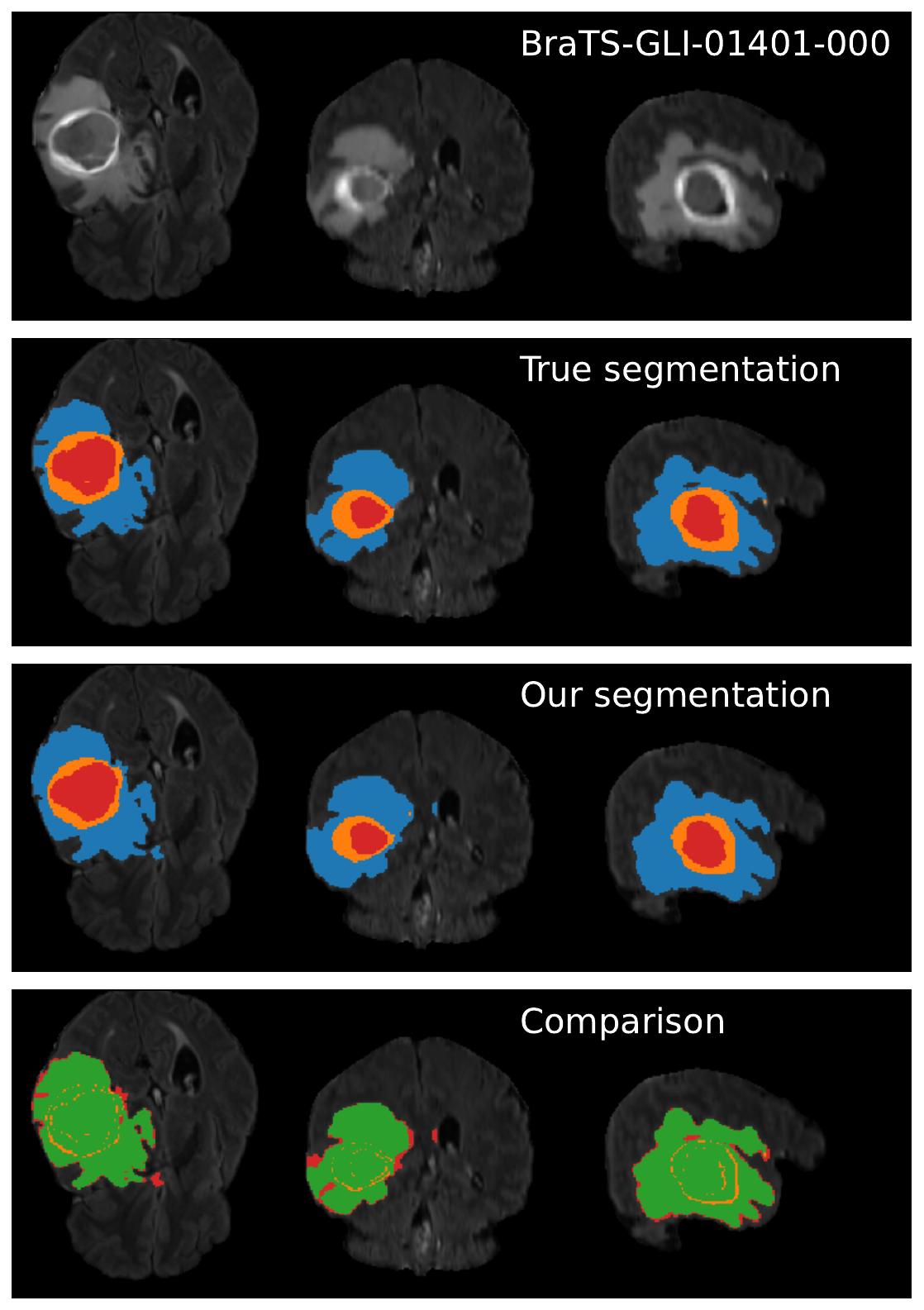}
			\includegraphics[width=.329\linewidth]{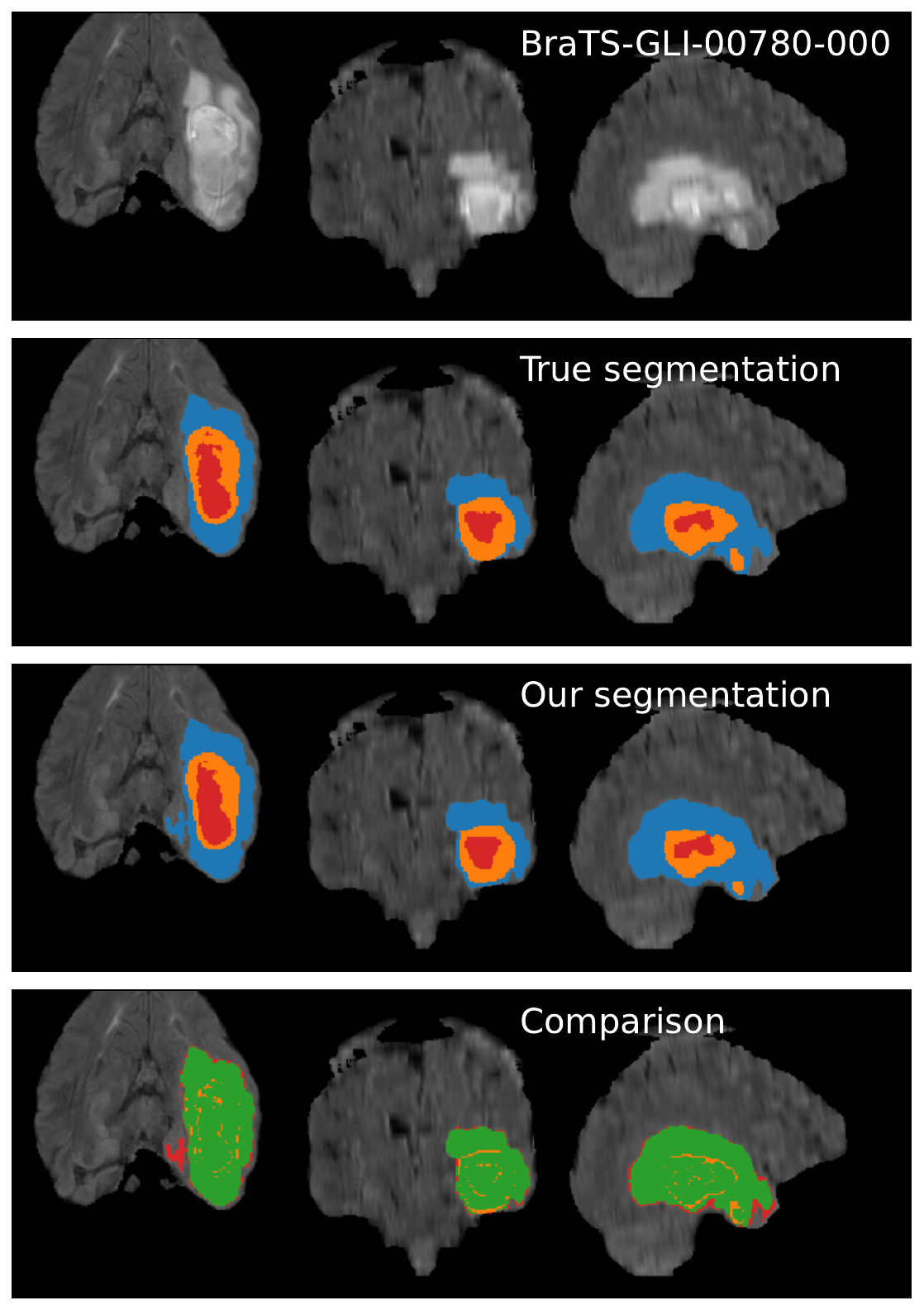}
			\includegraphics[width=.329\linewidth]{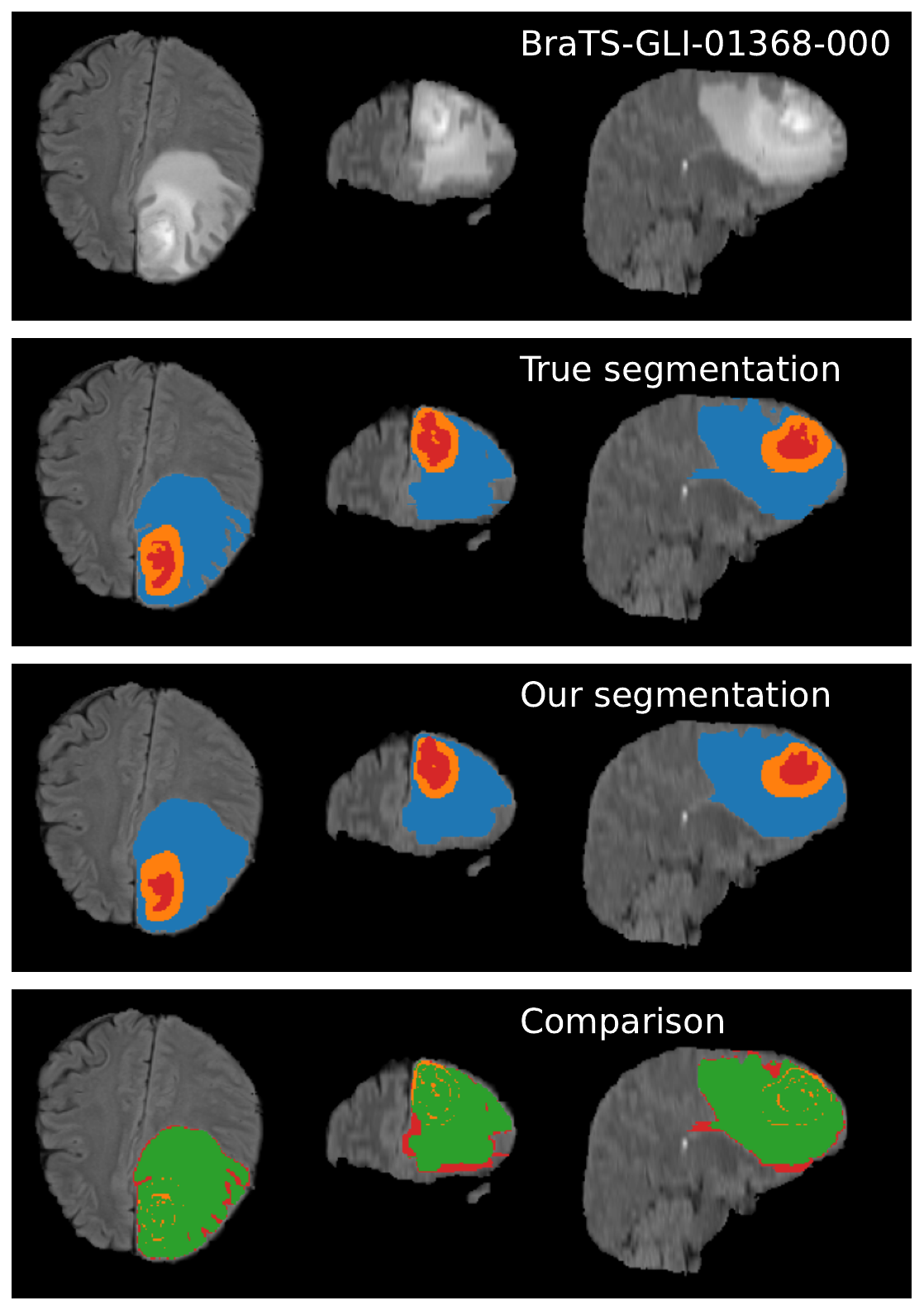}
		\end{minipage}
		\label{fig:examples_seg_good}}\\
	\subfloat[Cases where the model is not valid]{
		\begin{minipage}{.985\linewidth}
			\includegraphics[width=.329\linewidth]{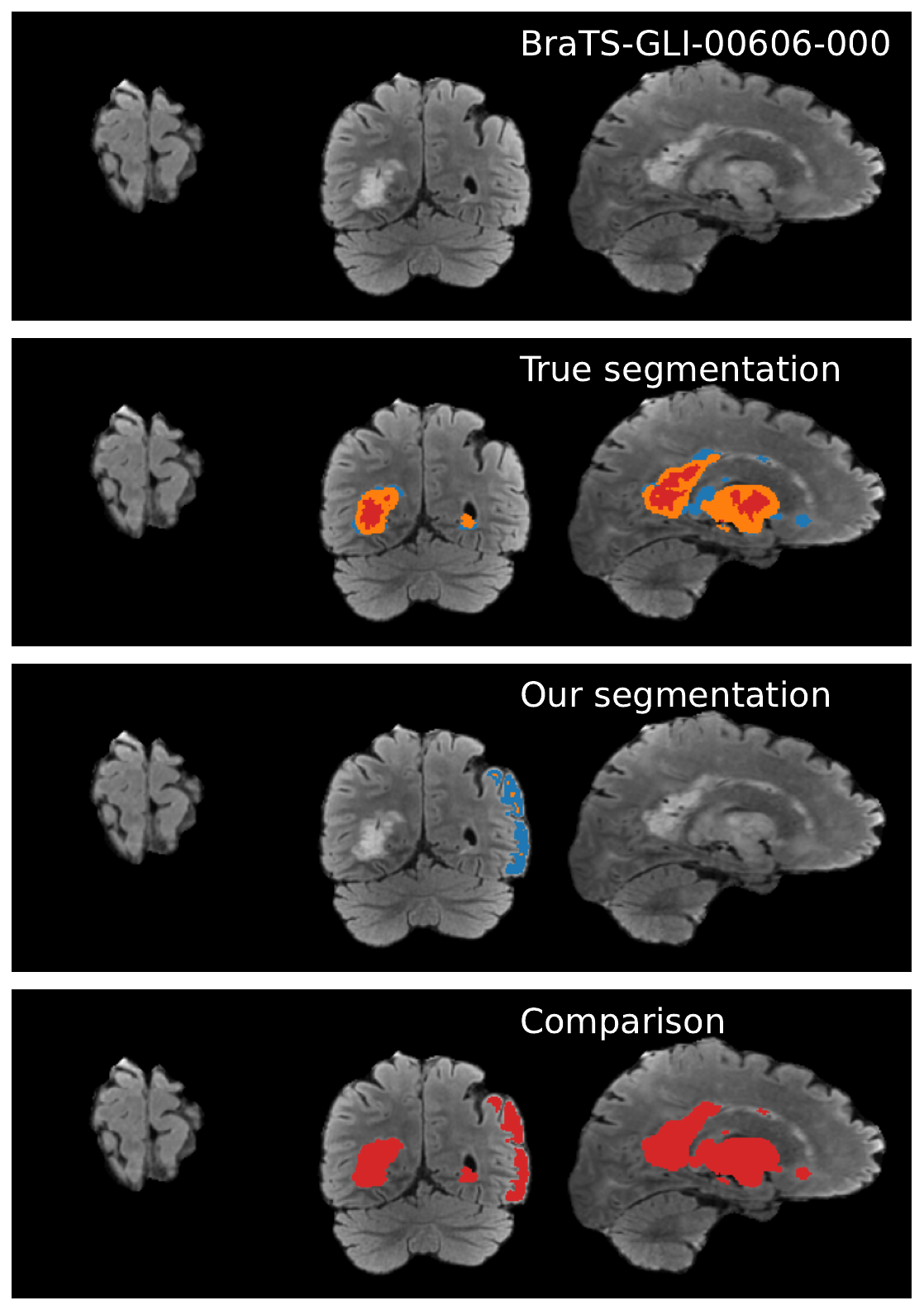}
			\includegraphics[width=.329\linewidth]{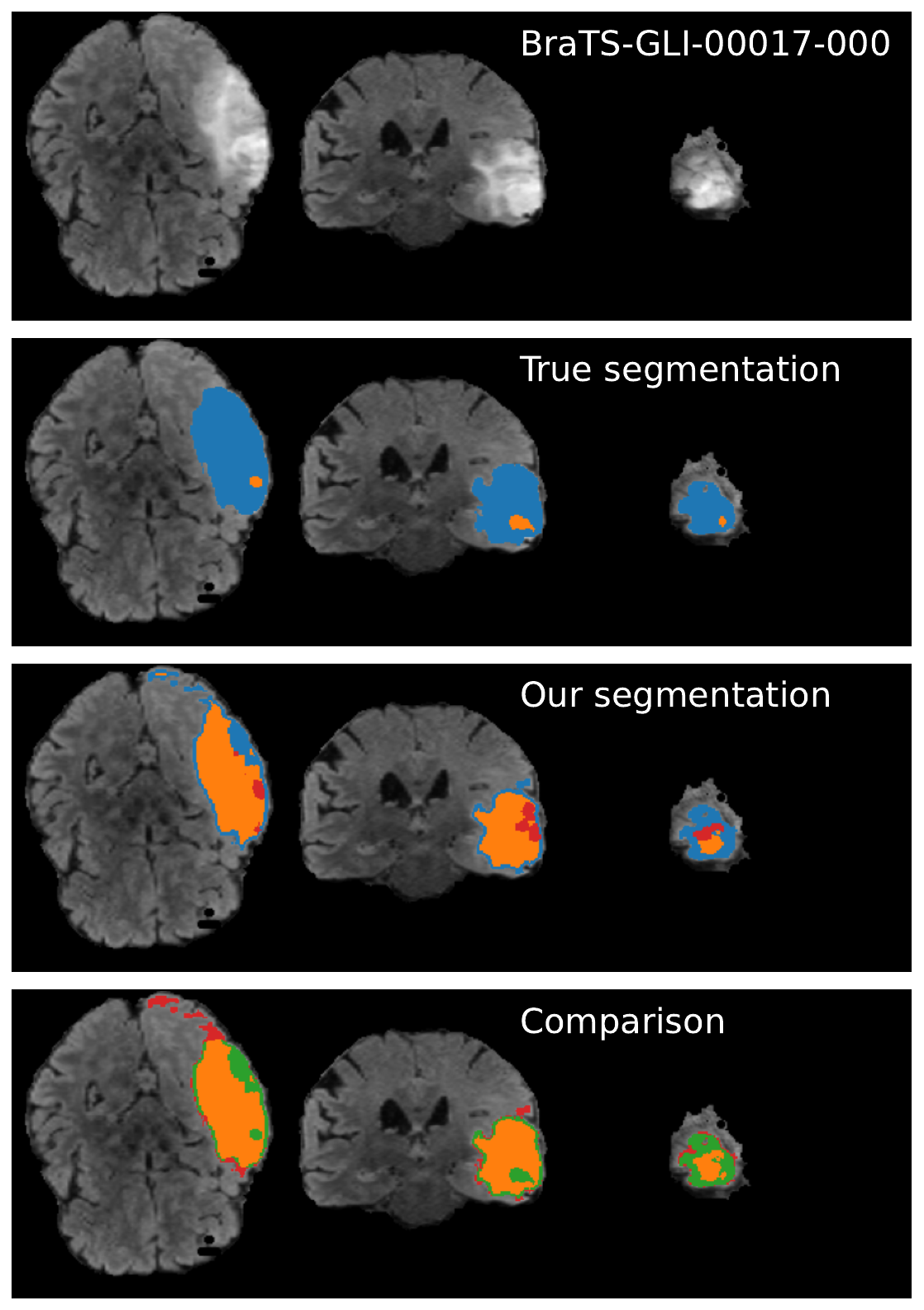}
			\includegraphics[width=.329\linewidth]{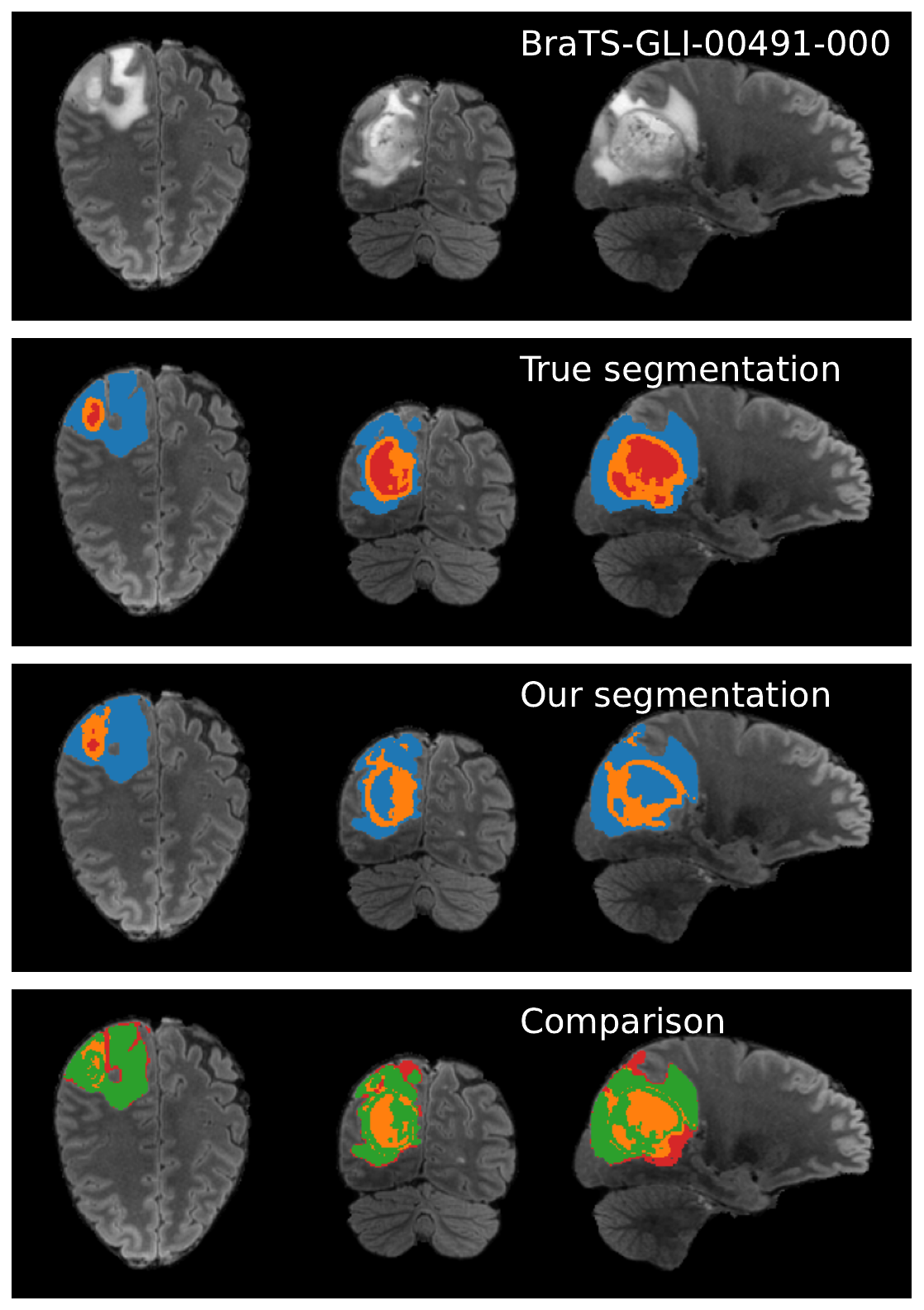}
		\end{minipage}
		\label{fig:examples_seg_bad}}
	\caption{Results of our algorithm on BraTS 2025, in cases where the model is valid or not. Each panel represents an image of modality FLAIR, the segmentation provided by the experts, the one obtained by our method (TC: red, ED: blue, ET: orange), and a comparison (green: correctly segmented, orange: mislabeled, red: false positive or negative).}
\end{figure*}

Inspecting low-performing cases (\Cref{fig:examples_seg_bad}) suggests that failures are often associated with violations of the applicability conditions from \Cref{sec:model}.
We observed that the enhancing tumor does not surround the necrosis, or only partially, forming a perforated sphere. 
In both cases, the algorithm cannot partition the domain into the interior and exterior of ET, leading to an incorrect estimation of the other components.
We analyze these failure modes qualitatively in \Cref{subsec:qualitative_eval}.

There is a notable discrepancy between the mean and median values, which are approximately 0.88 (WT), 0.16 (TC), 0.62 (ED), and 0.52 (ET).
Apart from TC, the median values are all higher. 
This difference reflects substantial variability, depending on whether the model hypotheses are satisfied, as examined next.

\paragraph{Importance of the topological model}

In order to evaluate the importance of the model, we restricted the analysis to only a subset of images, those where hypotheses \ref{ch4:enum:global} to \ref{ch4:enum:TC}, described in \Cref{sec:model}, are valid.
This represents 441 images (35.3\% of the dataset).
The Dice scores restricted to this subset of images are presented in \Cref{fig:benchmark_all} (blue rectangles).
In this case, our method exhibits clear gains: the mean scores are $0.87\pm0.15$ (WT), $0.56\pm0.34$ (TC), $0.72\pm0.21$ (ED), and $0.64\pm0.22$ (ET).
These values respectively increased by $0.11$, $0.21$, $0.15$, and $0.18$.

\subsection{Qualitative evaluation and limitations}\label{subsec:qualitative_eval}

Overall, most outputs of our method are biologically plausible, as illustrated in \Cref{fig:examples_seg_good}. In the third row, the method captures fine details of the tumor outline, and in all three examples all labels are correctly identified. 

Several situations can cause our algorithm to fail, and all of them fall outside the scope of our hypotheses (see examples in \Cref{fig:examples_seg_bad}):
\begin{itemize}
	\item \textbf{WT not hyper-intense in FLAIR} (see left panel).
	If the tumor does not stand out as a bright foreground in FLAIR, \ref{item:step1} may miss large parts of WT, which propagates to all subsequent labels and yields a poor final segmentation.
	\item \textbf{Incorrect topological signature for ET} (see middle panel).
	Even when WT is correctly detected, the bright structure in T1ce may not exhibit the expected topology (e.g., it is a solid ball rather than spherical), in which case \ref{item:step2} may select an incorrect $H_2$ feature and misidentify ET.
	\item \textbf{ET does not separate inside/outside} (see right panel).
	In some cases, ET is detected but the resulting segmentation is a perforated sphere, lacking a well-defined interior and exterior.  
	This causes \ref{item:step3} to misclassify TC versus ED.
\end{itemize}

\subsection{Comparison with other methods}\label{subsubsec:brain_comparison}

We compared our method with AUCseg \cite{zhaoAUCseg2021}, another unsupervised segmentation framework based on clustering and morphological processing. 
To our knowledge, it is the only unsupervised method with public implementations.
AUCseg proceeds in three stages, each exploiting MRI contrasts to delineate tumor subregions. 
First, WT is segmented via k-means clustering on FLAIR images, where ED presents as high-intensity regions. 
Next, using the WT mask as a region of interest, clustering is applied to T1ce images to identify ET.
Finally, necrotic areas are extracted through morphological operations or, when ET is not fully connected, via clustering on T2 images. 
From this point of view, their pipeline is similar to ours.

In their original work, the authors of AUCseg reported competitive results on the BraTS 2018 dataset, with Dice scores of approximately 0.82 (WT), 0.71 (TC), and 0.73 (ET).
However, the results we obtained on BraTS 2025, presented in \Cref{table:brains_comparison}, were lower. 
While AUCseg provides tunable parameters, we limited evaluation to two pipelines proposed by the authors: one assuming ET encloses TC (\textbf{AUCseg:CC}) and another one that does not (\textbf{AUCseg:T2}). 
We kept the number of k-means clusters at the recommended default values.

Our method achieves higher Dice scores for WT, TC, and ED. 
The improvements are especially pronounced for WT (+0.1 absolute Dice, approximately 15.2\% relative improvement) and TC (+0.07 absolute Dice, 25.9\% improvement). 
For ET, AUCseg's mean score is higher (+0.08).

Regarding DL methods, in BraTS 2023, U-Nets achieved Dice scores of $0.846$ (ET), $0.876$ (TC), and $0.929$ (WT) \cite{ferreira2024wonbrats2023adult}. 
More recently, the top-performing approach in BraTS 2025 GLI-PRE \cite{parida2025improving} appears to rely on a similar strong ensemble strategy and reports validation Dice scores of $0.794$ (ET), $0.798$ (TC), and $0.881$ (WT). 
As expected, our train-free approach achieves lower performance than these fully supervised deep-learning models, whose optimization directly leverages large annotated training sets. 
We stress that these results are not directly comparable to ours: they are computed on different evaluation subsets, and the validation phase data are no longer accessible.

%
%

\section{Application to cortical plate segmentation}\label{sec:fetal}

\begin{figure*}[h!]
	\centering
	\includegraphics[width=\linewidth]{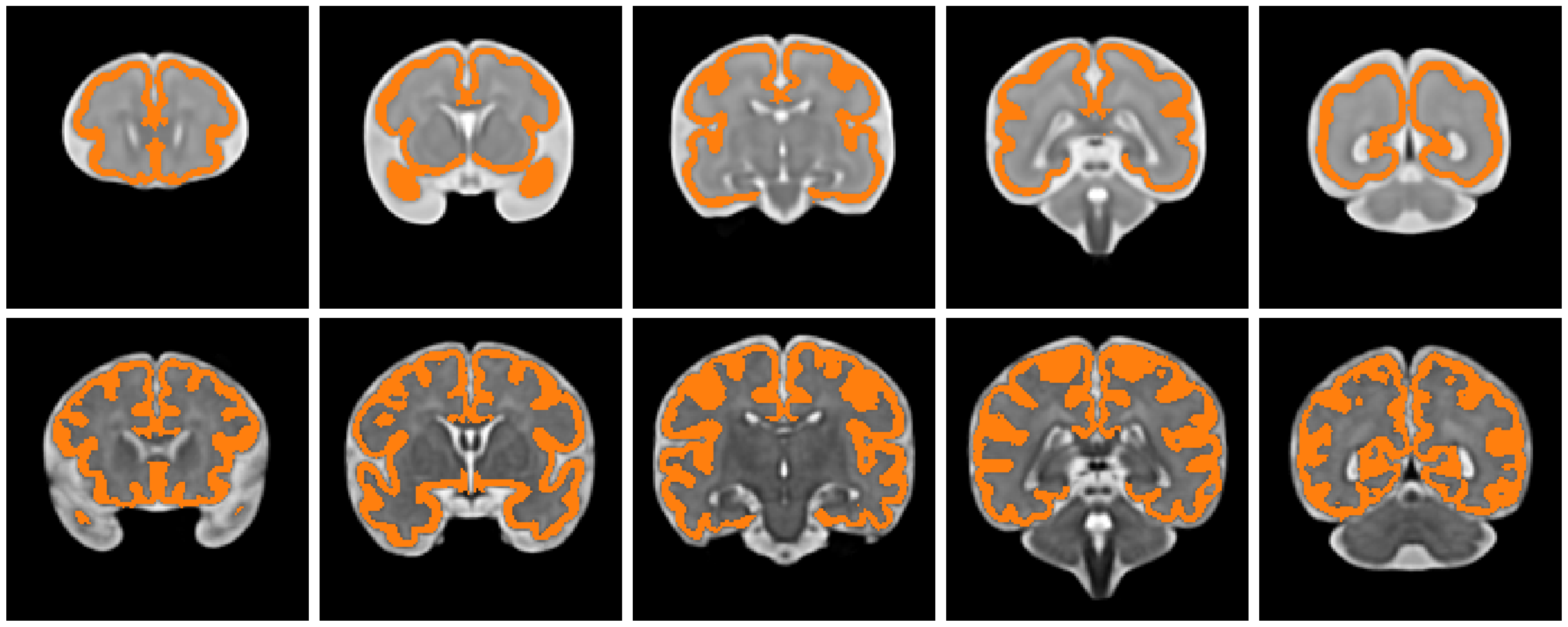}
	\caption{
		\textbf{Cortical plate segmentation in STA.}
		The rows contain coronal slices of the same MRI, for gestational week 30 (top) and 38 (bottom), with ground-truth cortical plate segmentation (orange).
	}
	\label{fig:fetal_coronal_slices}
\end{figure*}

The ideas developed in this article extend beyond glioma segmentation to other imaging tasks under comparable topological assumptions. 
We illustrate this through cortical plate (CP) segmentation in fetal brain MRI.

\subsection{Dataset}

Several neurological deficits---such as ventriculomegaly, which is associated with schizophrenia, autism and epilepsy---can be detected during fetal development \cite{makropoulos2018review}.
In particular, the correct maturation of the fetal brain can be observed through the \textit{gyrification} of the cortical plate, the embryonic precursor of the cerebral cortex.
From the tenth to the thirty-fifth gestational week, the CP changes
from a smooth surface to a highly convoluted one, making its segmentation a difficult task \cite{ciceri2023review}.

To date, only a few datasets are publicly available \cite{ciceri2024fetal}.
The first one was the \textit{Fetal Tissue Annotation and Segmentation Dataset} (FeTA) \cite{payette2021automatic,payette2023fetal}, led by the University Children's Hospital Zürich and the University of Zürich, which gathered 50 manually segmented pathological and non-pathological fetal brain MRIs,
across a range of gestational ages (20 to 33 weeks).
On the other hand, a number of atlases are found online, such as the \textit{Spatiotemporal Atlas} (STA) \cite{gholipour2017normative}.
It contains representations of the \textit{average} fetal brain, at one-week intervals between 21 and 38 weeks gestational age (\Cref{fig:fetal_coronal_slices}).
These images were obtained via diffeomorphic deformable registration of 81 T2-weighted MRI scans of healthy fetuses.
Compared with the clinical FeTA dataset, atlas data are smoother and easier to process, and are therefore used in this work.
We applied a simple preprocessing to each image: a 0--1 normalization and a Gaussian blur with standard deviation $\sigma=0.5$.

\subsection{Framework in 2D}

Because the cortical plate is the only class to identify, we only need \ref{item:step2} of our method.
However, as shown in \Cref{fig:fetal_coronal_slices}, it forms a perforated sphere, open at the cerebellar level (third and fourth slices of both rows).
Consequently, the homology groups of the cortical plate are all trivial, and it cannot be detected with our method.
We consider a different strategy: we study the image slice by slice, in the coronal plane.

We observed that most slices fall into one of three types: the cortical plate forms either (i) one circle (first two slices in both rows of \Cref{fig:fetal_coronal_slices}), (ii) two disjoint circles (last slice), or (iii) an ``open circle'', i.e., a circular arc (third and fourth slices).
These cases are identified by inspecting the persistence diagram as follows.

\begin{enumerate}
\item 
	We compute the $H_1$-persistence of the superlevel set filtration of the slice.
	For each point in the persistence diagram, we identify the associated component, as in \ref{item:step2} (connected component of the birth pixel at birth time).

\item 
	We discard points in the diagram that lead to implausible segmentations.
	Concretely, for each candidate component, we measure the area of the enclosed hole and retain the candidate only if this area lies within $[N_\text{min},N_\text{max}]$.
	In practice, we set $N_\text{min}$ and $N_\text{max}$ to 25\% and 75\% of the slice area, respectively.
\item 
	Among the remaining candidates, we select an ``optimal feature'' according to the criterion defined below.
	We classify a slice as type (ii) if there exists another point of the diagram at distance at most $\epsilon=0.1$ (a fixed threshold), and type (i) otherwise.
		For simplicity, slices of type (iii) are treated as type (i).
\end{enumerate}
The final segmentation is defined as the union of the connected components associated with the selected points---one point for types (i) and (iii), and two points for type (ii).

To select the ``optimal feature'', i.e., that corresponding to the CP, we considered three criteria: 
\begin{itemize}
	\item 
		\textbf{Earliest-born:} 
		the point with minimal birth time, favoring promptly appearing cycles.
	\item 
		\textbf{Largest area:} 
		the point whose associated connected component encloses the most pixels, favoring ``larger cycles''.
	\item 
		\textbf{Most persistent point:} 
		the point maximizing the persistence $(\text{death}-\text{birth})$, favoring features most robust to noise.
\end{itemize}
In practice, the three selectors produced comparable results. 
The second option---choosing the loop that encloses the largest area---performed slightly better, so we adopt it as our default.

\subsection{Scores}
\begin{figure}[h!]
	\centering
	\includegraphics[width=\linewidth]{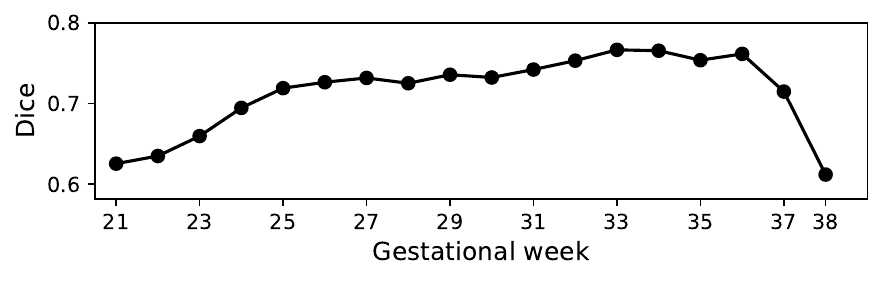}
	\caption{
		\textbf{Results of our method on STA.}
		Evolution of Dice scores obtained for cortical plate segmentation on STA, as a function of the gestational week.
		The min, max, and average values are 0.611 (week 38), 0.767 (week 33), and 0.714.
	}
	\label{fig:fetal_scores_week}
\end{figure}

Over the full STA collection of 18 images, we obtained a mean Dice score of 0.714 (standard deviation 0.048) against the ground-truth segmentation.
The scores are displayed in \Cref{fig:fetal_scores_week} as a function of the gestational age (weeks) and are also presented in \Cref{table:fetal_comparison}, where they will be used for comparison to other methods.

The score declines over gestational age, as expected, because the cortical plate becomes increasingly folded and therefore harder to segment.
This phenomenon is also observed with TopoCP \cite[Fig.\ 11]{de2022multi}, where the Dice score drops from approximately 0.825 at gestational week 25 to 0.7 at week 38.

\begin{table*}[h!]
	\centering
	\begin{tabular}{l|cccc}
		\toprule
		\diagbox{\textbf{Region}}{\textbf{Method}} & \textbf{TDA} & \textbf{U-Net} & \textbf{TopoCP'21} & \textbf{TopoCP'22} \\
		\midrule
		CP & $0.71\pm0.05$ & $0.54\pm0.16$ & $0.70\pm0.14$ & \textbf{0.79$\pm$0.05} \\
		\bottomrule
	\end{tabular}
	\caption{
		\textbf{Comparison with other methods on STA.}
		Mean Dice score and standard deviation for cortical plate (CP) segmentation on the STA dataset, obtained by our method (\textbf{TDA}), a baseline \textbf{U-Net} and \textbf{TopoCP'21} in \cite[Table 2]{de2021segmentation}, and \textbf{TopoCP'22} in \cite[Table 4]{de2022multi}.
		The highest score is shown in bold.
	}	
	\label{table:fetal_comparison}
\end{table*}

\subsection{Validation of the model}\label{subsubsec:fetal_model}

Since our fetal brain segmentation task involves only one class---the cortical plate---only the second hypothesis of our model in \Cref{sec:model} is relevant.
Remembering that, in coronal slices, the cortical plate is expected to form one or two circles, we propose the following reformulation:
\begin{itemize}
\item[\textbf{(H2'')}] 
	The cortical plate divides each cortical slice into several connected components. 
	After removing the background and the components of cardinality lower than a hundredth of the slice, one or two components remain.
\end{itemize}

We found that, averaged over the whole collection of 18 images, this assumption is valid on 56.95\% of the non-empty cortical slices.
In fact, two phenomena can cause the assumption to break: either the cortical plate consists of several convex connected components, or it forms an open circle (as observed in the middle slices of \Cref{fig:fetal_coronal_slices}).
Although not studied further in this work, these slices warrant special attention to improve the scores.

\subsection{Comparison with other methods}\label{subsubsec:fetal_comparison}

In 2021, Dumast et al.\ proposed TopoCP in \cite{de2021segmentation}, an automatic method for cortical plate segmentation, augmenting a 2D U-Net training with persistent-homology-based topological loss.
It is trained on FeTA data and evaluated on STA, reporting consistent gains over a baseline U-Net.
In a follow-up 2022 preprint \cite{de2022multi}, the same authors substantially strengthen this framework: they generalize the topological loss to a multi-dimensional formulation (combining $0$- and $1$-dimensional homology, with a weighted loss), and pair it with a more robust inference pipeline based on a 2.5D multi-view, together with data augmentation and an ensemble majority vote across cross-validation models.	
This led to notable improvements in both classical metrics (Dice score) and topological metrics (Betti number error and hole ratio).
Their scores are reported in \Cref{table:fetal_comparison}, together with those of our method.

The table shows that TopoCP'22 attains the highest mean Dice on STA (0.79), followed by our method (0.71), TopoCP'21 (0.70), and finally U-Net (0.54).
Our method improves substantially over the baseline U-Net and is on par with TopoCP'21 on this dataset.
It is noteworthy that our scores were obtained without any training, whereas the other methods were trained on FeTA.

%
%

\section{Conclusion}\label{sec:conclusions}

This work investigated what \emph{train-free} topological data analysis can contribute to medical image segmentation when used as a primary inference mechanism rather than as an auxiliary signal for learning.
We proposed a simple, modular pipeline that combines classical intensity-based operations with persistent-homology cues.
Importantly, the framework remains interpretable end-to-end.

On BraTS 2025, our method yields robust Whole Tumor masks and competitive scores relative to a publicly available unsupervised baseline (AUCseg).
Although it does not reach the performance of deep-learning methods, it is well suited to scarce-data settings (when labels are costly or unavailable), scenarios where explainability and topological guarantees are important, and it can serve as a warm start for expert editing or for initializing and regularizing network training.

We further illustrated how the same framework can be adapted to a different anatomical target in fetal brain MRI by moving to a 2D formulation for cortical plate segmentation on STA.
Remarkably, in this setting, our train-free approach improves over a baseline U-Net and matches TopoCP'21, while remaining below TopoCP'22.

Crucially, the performance of our approach is strongly conditioned by the validity of the underlying intensity and topology assumptions: when the applicability conditions of our topological model hold, Dice scores increase markedly across all tumor components.
In other words, the method can be a useful segmentation prior, but only within an explicit scope of applicability.

Promising directions include automatic fallback strategies when the model fails, more faithful representative-cycle extraction, and hybrid pipelines where topology-driven outputs serve as initialization, quality control, or regularization for learning-based methods.

\bibliography{main}

@ARTICLE{BRATS2021,
   author={Menze, Bjoern H. and others},
   journal={IEEE Transactions on Medical Imaging}, 
   title={The Multimodal Brain Tumor Image Segmentation Benchmark ({BraTS})}, 
   year ={2015},
   volume={34},
   number={10},
   pages={1993-2024},
   }

@article{AutomaticMeasurementSister,
title={Automatic measurement of sister chromatid exchange frequency.},
author={Zack, Gregory W and Rogers, William E and Latt, Samuel A},
journal={Journal of Histochemistry \& Cytochemistry},
volume={25},
number={7},
pages={741--753},
year={1977},
publisher={SAGE Publications Sage CA: Los Angeles, CA}
}

@inproceedings{carrSimplifyingFlexibleIsosurfaces2004,
	title={Simplifying flexible isosurfaces using local geometric measures},
	author={Carr, Hamish and Snoeyink, Jack and Van De Panne, Michiel},
	booktitle={IEEE Visualization 2004},
	pages={497--504},
	year={2004},
	organization={IEEE}
}

@article{tierny2017topology,
	title={The topology toolkit},
	author={Tierny, Julien and Favelier, Guillaume and Levine, Joshua A and Gueunet, Charles and Michaux, Michael},
	journal={IEEE transactions on visualization and computer graphics},
	volume={24},
	number={1},
	pages={832--842},
	year={2017},
	publisher={IEEE}
}

@article{ballesterTreeShapesImage2003,
	title={The tree of shapes of an image},
	author={Ballester, Coloma and Caselles, Vicent and Monasse, Pascal},
	journal={ESAIM: Control, Optimisation and Calculus of Variations},
	volume={9},
	pages={1--18},
	year={2003},
	publisher={EDP Sciences}
}

@article{carlinet2014comparative,
	title={A comparative review of component tree computation algorithms},
	author={Carlinet, Edwin and G{\'e}raud, Thierry},
	journal={IEEE Transactions on Image Processing},
	volume={23},
	number={9},
	pages={3885--3895},
	year={2014},
	publisher={IEEE}
}

@article{xuHierarchicalSegmentationUsing2017, 
 	title={Hierarchical Segmentation Using Tree-Based Shape Spaces}, 
 	volume={39}, 
 	ISSN={2160-9292}, 
 	number={3}, 
 	journal={IEEE Transactions on Pattern Analysis and Machine Intelligence}, 
 	publisher={Institute of Electrical and Electronics Engineers (IEEE)}, 
 	author={Xu, Yongchao and Carlinet, Edwin and Geraud, Thierry and Najman, Laurent}, 
 	year={2017}, 
 	month=Mar, 
 	pages={457–469} }

@ARTICLE{Clough2020,
   author={Clough, James and Byrne, Nicholas and Oksuz, Ilkay and Zimmer, Veronika A. and Schnabel, Julia A. and King, Andrew},
   journal={IEEE TPAMI}, 
   title={A Topological Loss Function for Deep-Learning based Image Segmentation using Persistent Homology}, 
   year={2020}
   }

@article{zhang20213d,
	title={3D brain glioma segmentation in MRI through integrating multiple densely connected 2D convolutional neural networks},
	author={Zhang, Xiaobing and Hu, Yin and Chen, Wen and Huang, Gang and Nie, Shengdong},
	journal={Journal of Zhejiang University-SCIENCE B},
	volume={22},
	number={6},
	pages={462--475},
	year={2021},
	publisher={Springer}
}

@article{islam2020detection,
	title={Detection and classification of brain tumor based on multilevel segmentation with convolutional neural network},
	author={Islam, Rafiqul and Imran, Shah and Ashikuzzaman, Md and Khan, Md Munim Ali},
	journal={Journal of Biomedical Science and Engineering},
	volume={13},
	number={4},
	pages={45--53},
	year={2020},
	publisher={Scientific Research Publishing}
}

@article{liu2021canet,
	title={{CANet}: Context aware network for brain glioma segmentation},
	author={Liu, Zhihua and Tong, Lei and Chen, Long and Zhou, Feixiang and Jiang, Zheheng and Zhang, Qianni and Wang, Yinhai and Shan, Caifeng and Li, Ling and Zhou, Huiyu},
	journal={IEEE Transactions on Medical Imaging},
	volume={40},
	number={7},
	pages={1763--1777},
	year={2021},
	publisher={IEEE}
}

@article{zhaoAUCseg2021,
	title={{AUCseg}: An automatically unsupervised clustering toolbox for 3d-segmentation of high-grade gliomas in multi-parametric mr images},
	author={Zhao, Botao and Ren, Yan and Yu, Ziqi and Yu, Jinhua and Peng, Tingying and Zhang, Xiao-Yong},
	journal={Frontiers in Oncology},
	volume={11},
	pages={679952},
	year={2021}
}

@article{brats_2021,
  author    = {Ujjwal, Baid and others},
  title     = {The {RSNA-ASNR-MICCAI} {BraTS} 2021 Benchmark on Brain Tumor Segmentation and Radiogenomic Classification},
  journal   = {arXiv:2107.02314},
  year      = {2021}
}

@article{Bakas2017,
  year = {2017},
  volume = {4},
  number = {1},
  author = {Spyridon Bakas and Hamed Akbari and Aristeidis Sotiras and Michel Bilello and Martin Rozycki and Justin S. Kirby and John B. Freymann and Keyvan Farahani and Christos Davatzikos},
  title = {Advancing The Cancer Genome Atlas glioma {MRI} collections with expert segmentation labels and radiomic features},
  journal = {Scientific Data}
}

@inproceedings{luu2022extending,
	title={Extending nn-UNet for brain tumor segmentation},
	author={Luu, Huan Minh and Park, Sung-Hong},
	booktitle={International MICCAI Brainlesion Workshop},
	pages={173--186},
	year={2022},
	organization={Springer}
}

@inproceedings{yuan2022evaluating,
	title={Evaluating Scale Attention Network for Automatic Brain Tumor Segmentation with Large Multi-parametric MRI Database},
	author={Yuan, Yading},
	booktitle={International MICCAI Brainlesion Workshop},
	pages={42--53},
	year={2022},
	organization={Springer}
}

@inproceedings{futrega2022optimized,
	title={Optimized {U-Net} for brain tumor segmentation},
	author={Futrega, Micha{\l} and Milesi, Alexandre and Marcinkiewicz, Micha{\l} and Ribalta, Pablo},
	booktitle={International MICCAI Brainlesion Workshop},
	pages={15--29},
	year={2022},
	organization={Springer}
}

@inproceedings{siddiquee2021redundancy,
	title={Redundancy reduction in semantic segmentation of 3d brain tumor mris},
	author={Rahman Siddiquee, Md Mahfuzur and Myronenko, Andriy},
	booktitle={International MICCAI Brainlesion Workshop},
	pages={163--172},
	year={2021},
	organization={Springer}
}

@inproceedings{ma2022nnunet,
	title={NnUNet with Region-based Training and Loss Ensembles for Brain Tumor Segmentation},
	author={Ma, Jun and Chen, Jianan},
	booktitle={International MICCAI Brainlesion Workshop},
	pages={421--430},
	year={2022},
	organization={Springer}
}

@inproceedings{kotowski2022coupling,
	title={Coupling nn{U}-{N}ets with Expert Knowledge for Accurate Brain Tumor Segmentation from MRI},
	author={Kotowski, Krzysztof and Adamski, Szymon and Machura, Bartosz and Zarudzki, Lukasz and Nalepa, Jakub},
	booktitle={International MICCAI Brainlesion Workshop},
	pages={197--209},
	year={2022},
	organization={Springer}
}

@inproceedings{ren2022ensemble,
	title={Ensemble Outperforms Single Models in Brain Tumor Segmentation},
	author={Ren, Jianxun and Zhang, Wei and An, Ning and Hu, Qingyu and Zhang, Youjia and Zhou, Ying},
	booktitle={International MICCAI Brainlesion Workshop},
	pages={451--462},
	year={2022},
	organization={Springer}
}

@inproceedings{jia2022hnf,
	title={HNF-Netv2 for brain tumor segmentation using multi-modal MR imaging},
	author={Jia, Haozhe and Bai, Chao and Cai, Weidong and Huang, Heng and Xia, Yong},
	booktitle={International MICCAI Brainlesion Workshop},
	pages={106--115},
	year={2021},
	organization={Springer}
}

@misc{ferreira2024wonbrats2023adult,
	title={How we won BraTS 2023 Adult Glioma challenge? Just faking it! Enhanced Synthetic Data Augmentation and Model Ensemble for brain tumour segmentation}, 
	author={André Ferreira and Naida Solak and Jianning Li and Philipp Dammann and Jens Kleesiek and Victor Alves and Jan Egger},
	year={2024},
	eprint={2402.17317},
	archivePrefix={arXiv},
	primaryClass={eess.IV},
}

@inproceedings{jiang2020twostagecascadedunet,
	title={Two-stage cascaded {U-Net}: 1st place solution to brats challenge 2019 segmentation task},
	author={Jiang, Zeyu and Ding, Changxing and Liu, Minfeng and Tao, Dacheng},
	booktitle={International MICCAI brainlesion workshop},
	pages={231--241},
	year={2019},
	organization={Springer}
}

@inproceedings{kofler2023approachingpeakgroundtruth,
	title={Approaching peak ground truth},
	author={Kofler, Florian and Wahle, Johannes and Ezhov, Ivan and Wagner, Sophia J and Al-Maskari, Rami and Gryska, Emilia and Todorov, Mihail and Bukas, Christina and Meissen, Felix and Peng, Tingying and others},
	booktitle={2023 IEEE 20th International Symposium on Biomedical Imaging (ISBI)},
	pages={1--6},
	year={2023},
	organization={IEEE}
}

@article{parida2025improving,
	title={Improving Pre-trained Segmentation Models using Post-Processing},
	author={Parida, Abhijeet and Capell{\'a}n-Mart{\'\i}n, Daniel and Jiang, Zhifan and Kulkarni, Nishad and Iyer, Krithika and Tapp, Austin and Anwar, Syed Muhammad and Ledesma-Carbayo, Mar{\'\i}a J and Linguraru, Marius George},
	journal={arXiv preprint arXiv:2512.14937},
	year={2025}
}

@article{sri_template,
  year = {2010},
  volume = {31},
  number = {5},
  author = {Rohlfing Torsten and  Natalie M. Zahr and  Edith V. Sullivan and  Adolf Pfefferbaum},
  title = {The {SRI24} multichannel atlas of normal adult human brain structure},
  journal = {Human brain mapping},
  pages = {798-819}
}

@article{carlsson2009topology,
  title={Topology and data},
  author={Carlsson, Gunnar},
  journal={Bulletin of the American Mathematical Society},
  volume={46},
  number={2},
  pages={255--308},
  year={2009}
}

@article{chazal2021introduction,
  title={An introduction to {T}opological {D}ata {A}nalysis: fundamental and practical aspects for data scientists},
  author={Chazal, Fr{\'e}d{\'e}ric and Michel, Bertrand},
  journal={Frontiers in Artificial Intelligence},
  volume={4},
  year={2021},
  publisher={Frontiers Media SA}
}

@incollection{oudot2017persistence,
  title={Persistence theory: from quiver representations to data analysis},
  author={Oudot, Steve Y},
  volume={209},
  year={2017},
  publisher={American Mathematical Soc.}
}

@article{rucco2020towards,
  title={Towards personalized diagnosis of glioblastoma in fluid-attenuated inversion recovery (FLAIR) by topological interpretable machine learning},
  author={Rucco, Matteo and Viticchi, Giovanna and Falsetti, Lorenzo},
  journal={Mathematics},
  volume={8},
  number={5},
  pages={770},
  year={2020},
  publisher={Multidisciplinary Digital Publishing Institute}
}

@article{crawford2020predicting,
  title={Predicting clinical outcomes in glioblastoma: an application of topological and functional data analysis},
  author={Crawford, Lorin and Monod, Anthea and Chen, Andrew X and Mukherjee, Sayan and Rabad{\'a}n, Ra{\'u}l},
  journal={Journal of the American Statistical Association},
  volume={115},
  number={531},
  pages={1139--1150},
  year={2020},
  publisher={Taylor \& Francis}
}

@article{tauzin2021giotto,
  title={giotto-tda: A Topological Data Analysis Toolkit for Machine Learning and Data Exploration.},
  author={Tauzin, Guillaume and Lupo, Umberto and Tunstall, Lewis and P{\'e}rez, Julian Burella and Caorsi, Matteo and Medina-Mardones, Anibal M and Dassatti, Alberto and Hess, Kathryn},
  journal={J. Mach. Learn. Res.},
  volume={22},
  number={39},
  pages={1--6},
  year={2021}
}

@article{kaji2020cubical,
  title={Cubical ripser: Software for computing persistent homology of image and volume data},
  author={Kaji, Shizuo and Sudo, Takeki and Ahara, Kazushi},
  journal={arXiv preprint arXiv:2005.12692},
  year={2020}
}

@inproceedings{maria2014gudhi,
  title={The gudhi library: Simplicial complexes and persistent homology},
  author={Maria, Cl{\'e}ment and Boissonnat, Jean-Daniel and Glisse, Marc and Yvinec, Mariette},
  booktitle={International congress on mathematical software},
  pages={167--174},
  year={2014},
  organization={Springer}
}

@inproceedings{clough2019explicit,
  title={Explicit topological priors for deep-learning based image segmentation using persistent homology},
  author={Clough, James R and Oksuz, Ilkay and Byrne, Nicholas and Schnabel, Julia A and King, Andrew P},
  booktitle={International Conference on Information Processing in Medical Imaging},
  pages={16--28},
  year={2019},
  organization={Springer}
}

@article{qaiser2019fast,
  title={Fast and accurate tumor segmentation of histology images using persistent homology and deep convolutional features},
  author={Qaiser, Talha and Tsang, Yee-Wah and Taniyama, Daiki and Sakamoto, Naoya and Nakane, Kazuaki and Epstein, David and Rajpoot, Nasir},
  journal={Medical image analysis},
  volume={55},
  pages={1--14},
  year={2019},
  publisher={Elsevier}
}

@article{qaiser2016persistent,
  title={Persistent homology for fast tumor segmentation in whole slide histology images},
  author={Qaiser, Talha and Sirinukunwattana, Korsuk and Nakane, Kazuaki and Tsang, Yee-Wah and Epstein, David and Rajpoot, Nasir},
  journal={Procedia Computer Science},
  volume={90},
  pages={119--124},
  year={2016},
  publisher={Elsevier}
}

@article{obayashi2018volume,
  title={Volume-optimal cycle: Tightest representative cycle of a generator in persistent homology},
  author={Obayashi, Ippei},
  journal={SIAM Journal on Applied Algebra and Geometry},
  volume={2},
  number={4},
  pages={508--534},
  year={2018},
  publisher={SIAM}
}

@inproceedings{escolar2016optimal,
  title={Optimal cycles for persistent homology via linear programming},
  author={Escolar, Emerson G and Hiraoka, Yasuaki},
  booktitle={Optimization in the Real World: Toward Solving Real-World Optimization Problems},
  pages={79--96},
  year={2016},
  organization={Springer}
}

@article{bakas2018identifying,
  title={Identifying the best machine learning algorithms for brain tumor segmentation, progression assessment, and overall survival prediction in the BRATS challenge},
  author={Bakas, Spyridon and Reyes, Mauricio and Jakab, Andras and Bauer, Stefan and Rempfler, Markus and Crimi, Alessandro and Shinohara, Russell Takeshi and Berger, Christoph and Ha, Sung Min and Rozycki, Martin and others},
  journal={arXiv preprint arXiv:1811.02629},
  year={2018}
}

@article{skaf2022topological,
  title={Topological data analysis in biomedicine: A review},
  author={Skaf, Yara and Laubenbacher, Reinhard},
  journal={Journal of Biomedical Informatics},
  volume={130},
  pages={104082},
  year={2022},
  publisher={Elsevier}
}

@article{singh2023topological,
  title={Topological data analysis in medical imaging: current state of the art},
  author={Singh, Yashbir and Farrelly, Colleen M and Hathaway, Quincy A and Leiner, Tim and Jagtap, Jaidip and Carlsson, Gunnar E and Erickson, Bradley J},
  journal={Insights into Imaging},
  volume={14},
  number={1},
  pages={1--10},
  year={2023},
  publisher={SpringerOpen}
}

@inproceedings{bauer2010atlas,
  title={Atlas-based segmentation of brain tumor images using a Markov random field-based tumor growth model and non-rigid registration},
  author={Bauer, Stefan and Seiler, Christof and Bardyn, Thibaut and Buechler, Philippe and Reyes, Mauricio},
  booktitle={2010 annual international conference of the IEEE engineering in medicine and biology},
  pages={4080--4083},
  year={2010},
  organization={IEEE}
}

@inproceedings{zikic2012decision,
  title={Decision forests for tissue-specific segmentation of high-grade gliomas in multi-channel MR},
  author={Zikic, Darko and Glocker, Ben and Konukoglu, Ender and Criminisi, Antonio and Demiralp, Cagatay and Shotton, Jamie and Thomas, Owen M and Das, Tilak and Jena, Raj and Price, Stephen J},
  booktitle={Medical Image Computing and Computer-Assisted Intervention--MICCAI 2012: 15th International Conference, Nice, France, October 1-5, 2012, Proceedings, Part III 15},
  pages={369--376},
  year={2012},
  organization={Springer}
}

@article{wu2014brain,
  title={Brain tumor detection and segmentation in a CRF (conditional random fields) framework with pixel-pairwise affinity and superpixel-level features},
  author={Wu, Wei and Chen, Albert YC and Zhao, Liang and Corso, Jason J},
  journal={International journal of computer assisted radiology and surgery},
  volume={9},
  pages={241--253},
  year={2014},
  publisher={Springer}
}

@article{de2022multi,
  title={Multi-dimensional topological loss for cortical plate segmentation in fetal brain MRI},
  author={de Dumast, Priscille and Kebiri, Hamza and Dunet, Vincent and Koob, M{\'e}riam and Cuadra, Meritxell Bach},
  journal={arXiv preprint arXiv:2208.07566},
  year={2022}
}

@inproceedings{de2021segmentation,
  title={Segmentation of the cortical plate in fetal brain MRI with a topological loss},
  author={de Dumast, Priscille and Kebiri, Hamza and Atat, Chirine and Dunet, Vincent and Koob, M{\'e}riam and Cuadra, Meritxell Bach},
  booktitle={Uncertainty for Safe Utilization of Machine Learning in Medical Imaging, and Perinatal Imaging, Placental and Preterm Image Analysis: 3rd International Workshop, UNSURE 2021, and 6th International Workshop, PIPPI 2021, Held in Conjunction with MICCAI 2021, Strasbourg, France, October 1, 2021, Proceedings 3},
  pages={200--209},
  year={2021},
  organization={Springer}
}

@article{rucco2019fast,
  title={Fast Glioblastoma Detection in Fluid-attenuated inversion recovery (FLAIR) images by Topological Explainable Automatic Machine Learning},
  author={Rucco, Matteo and Viticchi, Giovanna},
  journal={arXiv preprint arXiv:1912.08167},
  year={2019}
}

@article{payette2023fetal,
  title={Fetal brain tissue annotation and segmentation challenge results},
  author={Payette, Kelly and Li, Hongwei Bran and de Dumast, Priscille and Licandro, Roxane and Ji, Hui and Siddiquee, Md Mahfuzur Rahman and Xu, Daguang and Myronenko, Andriy and Liu, Hao and Pei, Yuchen and others},
  journal={Medical Image Analysis},
  volume={88},
  pages={102833},
  year={2023},
  publisher={Elsevier}
}

@article{gholipour2017normative,
  title={A normative spatiotemporal MRI atlas of the fetal brain for automatic segmentation and analysis of early brain growth},
  author={Gholipour, Ali and Rollins, Caitlin K and Velasco-Annis, Clemente and Ouaalam, Abdelhakim and Akhondi-Asl, Alireza and Afacan, Onur and Ortinau, Cynthia M and Clancy, Sean and Limperopoulos, Catherine and Yang, Edward and others},
  journal={Scientific reports},
  volume={7},
  number={1},
  pages={476},
  year={2017},
  publisher={Nature Publishing Group UK London}
}

@article{payette2021automatic,
  title={An automatic multi-tissue human fetal brain segmentation benchmark using the fetal tissue annotation dataset},
  author={Payette, Kelly and de Dumast, Priscille and Kebiri, Hamza and Ezhov, Ivan and Paetzold, Johannes C and Shit, Suprosanna and Iqbal, Asim and Khan, Romesa and Kottke, Raimund and Grehten, Patrice and others},
  journal={Scientific data},
  volume={8},
  number={1},
  pages={167},
  year={2021},
  publisher={Nature Publishing Group UK London}
}

@article{makropoulos2018review,
  title={A review on automatic fetal and neonatal brain MRI segmentation},
  author={Makropoulos, Antonios and Counsell, Serena J and Rueckert, Daniel},
  journal={NeuroImage},
  volume={170},
  pages={231--248},
  year={2018},
  publisher={Elsevier}
}

@article{hu2019topology,
  title={Topology-preserving deep image segmentation},
  author={Hu, Xiaoling and Li, Fuxin and Samaras, Dimitris and Chen, Chao},
  journal={Advances in neural information processing systems},
  volume={32},
  year={2019}
}

@inproceedings{dey2010optimal,
  title={Optimal homologous cycles, total unimodularity, and linear programming},
  author={Dey, Tamal K and Hirani, Anil N and Krishnamoorthy, Bala},
  booktitle={Proceedings of the forty-second ACM symposium on Theory of computing},
  pages={221--230},
  year={2010}
}

@article{chen2011hardness,
  title={Hardness results for homology localization},
  author={Chen, Chao and Freedman, Daniel},
  journal={Discrete \& Computational Geometry},
  volume={45},
  number={3},
  pages={425--448},
  year={2011},
  publisher={Springer}
}

@article{cohen2022lexicographic,
  title={Lexicographic optimal homologous chains and applications to point cloud triangulations},
  author={Cohen-Steiner, David and Lieutier, Andr{\'e} and Vuillamy, Julien},
  journal={Discrete \& Computational Geometry},
  volume={68},
  number={4},
  pages={1155--1174},
  year={2022},
  publisher={Springer}
}

@article{li2021minimal,
  title={Minimal cycle representatives in persistent homology using linear programming: an empirical study with user’s guide},
  author={Li, Lu and Thompson, Connor and Henselman-Petrusek, Gregory and Giusti, Chad and Ziegelmeier, Lori},
  journal={Frontiers in artificial intelligence},
  volume={4},
  pages={681117},
  year={2021},
  publisher={Frontiers Media SA}
}

@article{obayashi2022persistent,
  title={Persistent homology analysis for materials research and persistent homology software: HomCloud},
  author={Obayashi, Ippei and Nakamura, Takenobu and Hiraoka, Yasuaki},
  journal={journal of the physical society of japan},
  volume={91},
  number={9},
  pages={091013},
  year={2022},
  publisher={The Physical Society of Japan}
}

@inproceedings{dey2019persistent,
  title={Persistent 1-cycles: Definition, computation, and its application},
  author={Dey, Tamal K and Hou, Tao and Mandal, Sayan},
  booktitle={Computational Topology in Image Context: 7th International Workshop, CTIC 2019, M{\'a}laga, Spain, January 24-25, 2019, Proceedings 7},
  pages={123--136},
  year={2019},
  organization={Springer}
}

@article{ciceri2024fetal,
	title={Fetal brain MRI atlases and datasets: A review},
	author={Ciceri, Tommaso and Casartelli, Luca and Montano, Florian and Conte, Stefania and Squarcina, Letizia and Bertoldo, Alessandra and Agarwal, Nivedita and Brambilla, Paolo and Peruzzo, Denis},
	journal={NeuroImage},
	volume={292},
	pages={120603},
	year={2024},
	publisher={Elsevier}
}

@article{gao2025medical,
	title={Medical Image Segmentation: A Comprehensive Review of Deep Learning-Based Methods},
	author={Gao, Yuxiao and Jiang, Yang and Peng, Yanhong and Yuan, Fujiang and Zhang, Xinyue and Wang, Jianfeng},
	journal={Tomography},
	volume={11},
	number={5},
	pages={52},
	year={2025},
	publisher={MDPI}
}

@article{verma2025brain,
	title={Brain tumor segmentation with deep learning: Current approaches and future perspectives},
	author={Verma, Akash and Yadav, Arun Kumar},
	journal={Journal of Neuroscience Methods},
	pages={110424},
	year={2025},
	publisher={Elsevier}
}

@article{sharma2010automated,
	title={Automated medical image segmentation techniques},
	author={Sharma, Neeraj and Aggarwal, Lalit M},
	journal={Journal of medical physics},
	volume={35},
	number={1},
	pages={3--14},
	year={2010},
	publisher={Medknow}
}

@article{pham2000current,
	title={Current methods in medical image segmentation},
	author={Pham, Dzung L and Xu, Chenyang and Prince, Jerry L},
	journal={Annual review of biomedical engineering},
	volume={2},
	number={1},
	pages={315--337},
	year={2000},
	publisher={Annual Reviews 4139 El Camino Way, PO Box 10139, Palo Alto, CA 94303-0139, USA}
}

@article{litjens2017survey,
	title={A survey on deep learning in medical image analysis},
	author={Litjens, Geert and Kooi, Thijs and Bejnordi, Babak Ehteshami and Setio, Arnaud Arindra Adiyoso and Ciompi, Francesco and Ghafoorian, Mohsen and Van Der Laak, Jeroen Awm and Van Ginneken, Bram and S{\'a}nchez, Clara I},
	journal={Medical image analysis},
	volume={42},
	pages={60--88},
	year={2017},
	publisher={Elsevier}
}

@article{ciceri2023review,
	title={Review on deep learning fetal brain segmentation from magnetic resonance images},
	author={Ciceri, Tommaso and Squarcina, Letizia and Giubergia, Alice and Bertoldo, Alessandra and Brambilla, Paolo and Peruzzo, Denis},
	journal={Artificial Intelligence in Medicine},
	volume={143},
	pages={102608},
	year={2023},
	publisher={Elsevier}
}

@article{bonato2025advancing,
	title={Advancing precision: A comprehensive review of MRI segmentation datasets from {BraTS} challenges (2012--2025)},
	author={Bonato, Beatrice and Nanni, Loris and Bertoldo, Alessandra},
	journal={Sensors (Basel, Switzerland)},
	volume={25},
	number={6},
	pages={1838},
	year={2025}
}

@article{de20242024,
	title={The 2024 brain tumor segmentation ({BraTS}) challenge: Glioma segmentation on post-treatment {MRI}},
	author={de Verdier, Maria Correia and Saluja, Rachit and Gagnon, Louis and LaBella, Dominic and Baid, Ujjwall and Tahon, Nourel Hoda and Foltyn-Dumitru, Martha and Zhang, Jikai and Alafif, Maram and Baig, Saif and others},
	journal={arXiv preprint arXiv:2405.18368},
	year={2024}
}

@article{nardi2024topology,
	title={Topology-based segmentation of 3D confocal images of emerging hematopoietic stem cells in the zebrafish embryo},
	author={Nardi, Giacomo and Torcq, L{\'e}a and Schmidt, AA and Olivo-Marin, J-C},
	journal={Biological Imaging},
	volume={4},
	pages={e11},
	year={2024},
	publisher={Cambridge University Press}
}

@article{panconi2024three,
	title={Three-dimensional topology-based analysis segments volumetric and spatiotemporal fluorescence microscopy},
	author={Panconi, Luca and Tansell, Amy and Collins, Alexander J and Makarova, Maria and Owen, Dylan M},
	journal={Biological Imaging},
	volume={4},
	pages={e1},
	year={2024},
	publisher={Cambridge University Press}
}

@inproceedings{gao2013segmenting,
	title={Segmenting the papillary muscles and the trabeculae from high resolution cardiac CT through restoration of topological handles},
	author={Gao, Mingchen and Chen, Chao and Zhang, Shaoting and Qian, Zhen and Metaxas, Dimitris and Axel, Leon},
	booktitle={International Conference on Information Processing in Medical Imaging},
	pages={184--195},
	year={2013},
	organization={Springer}
}

@inproceedings{chen2017cardiac,
	title={Cardiac trabeculae segmentation: an application of computational topology (multimedia contribution)},
	author={Chen, Chao and Metaxas, Dimitris and Wang, Yusu and Wu, Pengxiang},
	booktitle={33rd International Symposium on Computational Geometry (SoCG 2017)},
	pages={65--1},
	year={2017},
	organization={Schloss Dagstuhl--Leibniz-Zentrum f{\"u}r Informatik}
}

@article{faragallah2023efficient,
	title={Efficient brain tumor segmentation using OTSU and K-means clustering in homomorphic transform},
	author={Faragallah, Osama S and El-Hoseny, Heba M and El-sayed, Hala S},
	journal={Biomedical Signal Processing and Control},
	volume={84},
	pages={104712},
	year={2023},
	publisher={Elsevier}
}

@article{otsu1975threshold,
	author={Otsu, Nobuyuki},
	journal={IEEE Transactions on Systems, Man, and Cybernetics}, 
	title={A Threshold Selection Method from Gray-Level Histograms}, 
	year={1979},
	volume={9},
	number={1},
	pages={62-66}
}

@article{adams1994seeded,
	title={Seeded region growing},
	author={Adams, Rolf and Bischof, Leanne},
	journal={IEEE Transactions on pattern analysis and machine intelligence},
	volume={16},
	number={6},
	pages={641--647},
	year={1994},
	publisher={IEEE}
}

@inproceedings{mcqueen1967some,
	title={Some methods of classification and analysis of multivariate observations},
	author={McQueen, James B},
	booktitle={Proc. of 5th Berkeley Symposium on Math. Stat. and Prob.},
	pages={281--297},
	year={1967}
}

@article{dempster1977maximum,
	title={Maximum likelihood from incomplete data via the EM algorithm},
	author={Dempster, Arthur P and Laird, Nan M and Rubin, Donald B},
	journal={Journal of the royal statistical society: series B (methodological)},
	volume={39},
	number={1},
	pages={1--22},
	year={1977},
	publisher={Wiley Online Library}
}

@article{kass1988snakes,
	title={Snakes: Active contour models},
	author={Kass, Michael and Witkin, Andrew and Terzopoulos, Demetri},
	journal={International journal of computer vision},
	volume={1},
	number={4},
	pages={321--331},
	year={1988},
	publisher={Springer}
}

@article{chan2001active,
	title={Active contours without edges},
	author={Chan, Tony F and Vese, Luminita A},
	journal={IEEE Transactions on image processing},
	volume={10},
	number={2},
	pages={266--277},
	year={2001},
	publisher={IEEE}
}

@inproceedings{boykov2001interactive,
	title={Interactive graph cuts for optimal boundary \& region segmentation of objects in ND images},
	author={Boykov, Yuri Y and Jolly, M-P},
	booktitle={Proceedings eighth IEEE international conference on computer vision. ICCV 2001},
	volume={1},
	pages={105--112},
	year={2001},
	organization={IEEE}
}

@article{boykov2006graph,
	title={Graph cuts and efficient ND image segmentation},
	author={Boykov, Yuri and Funka-Lea, Gareth},
	journal={International journal of computer vision},
	volume={70},
	number={2},
	pages={109--131},
	year={2006},
	publisher={Springer}
}

@article{warfield2004simultaneous,
	title={Simultaneous truth and performance level estimation (STAPLE): an algorithm for the validation of image segmentation},
	author={Warfield, Simon K and Zou, Kelly H and Wells, William M},
	journal={IEEE transactions on medical imaging},
	volume={23},
	number={7},
	pages={903--921},
	year={2004},
	publisher={IEEE}
}

@article{iglesias2015multi,
	title={Multi-atlas segmentation of biomedical images: a survey},
	author={Iglesias, Juan Eugenio and Sabuncu, Mert R},
	journal={Medical image analysis},
	volume={24},
	number={1},
	pages={205--219},
	year={2015},
	publisher={Elsevier}
}

@article{cootes1995active,
	title={Active shape models-their training and application},
	author={Cootes, Timothy F and Taylor, Christopher J and Cooper, David H and Graham, Jim},
	journal={Computer vision and image understanding},
	volume={61},
	number={1},
	pages={38--59},
	year={1995},
	publisher={Elsevier}
}

@inproceedings{cootes1998active,
	title={Active appearance models},
	author={Cootes, Timothy F and Edwards, Gareth J and Taylor, Christopher J},
	booktitle={European conference on computer vision},
	pages={484--498},
	year={1998},
	organization={Springer}
}

@inproceedings{wu2017optimal,
	title={Optimal topological cycles and their application in cardiac trabeculae restoration},
	author={Wu, Pengxiang and Chen, Chao and Wang, Yusu and Zhang, Shaoting and Yuan, Changhe and Qian, Zhen and Metaxas, Dimitris and Axel, Leon},
	booktitle={International Conference on Information Processing in Medical Imaging},
	pages={80--92},
	year={2017},
	organization={Springer}
}

@misc{silvestre:hal-04871308,
	TITLE = {{Analyse dose-r{\'e}ponse {\`a} partir d'imagerie c{\'e}r{\'e}brale apr{\`e}s radioth{\'e}rapie via un m{\'e}lange d'expert spatial}},
	AUTHOR = {Silvestre, Th{\'e}o and Forbes, Florence and Ancelet, Sophie},
	NOTE = {Poster},
	HOWPUBLISHED = {{IABM 2024 - Colloque Fran{\c c}ais d'Intelligence Artificielle en Imagerie Biom{\'e}dicale}},
	ORGANIZATION = {{Instituts 3IA de Grenoble (MIAI), Nice (3IA C{\^o}te d'Azur) et Paris (PRAIRIE)}},
	PAGES = {1-1},
	YEAR = {2024},
	MONTH = Mar,
	HAL_ID = {hal-04871308},
	HAL_VERSION = {v1},
}

@article{wadell1935volume,
	title={Volume, shape, and roundness of quartz particles},
	author={Wadell, Hakon},
	journal={The Journal of geology},
	volume={43},
	number={3},
	pages={250--280},
	year={1935},
	publisher={University of Chicago Press}
}
\end{document}